\documentclass[12pt,preprint]{aastex}
\newcommand{\simgt}{\lower.5ex\hbox{$\; \buildrel > \over \sim \;$}}
\newcommand{\simlt}{\lower.5ex\hbox{$\; \buildrel < \over \sim \;$}}
\newcommand{\iint}{\int\!\!\!\int}
\newcommand{\bm}[1]{{\mbox{\boldmath${#1}$}}}
\newcommand{\gamps}{\gamma_{\rm ps}}
\newcommand{\gami}{\gamma_{\rm is}}
\newcommand{\gamo}{\gamma_{\rm os}}

\slugcomment{}
\shortauthors{Ohta, Taruya, \& Suto}
\shorttitle{Rings around Transiting Extrasolar Planets}

\begin{document}

\title{Predicting Photometric and Spectroscopic Signatures of Rings around
Transiting Extrasolar Planets}

\author{Yasuhiro Ohta, Atsushi Taruya\altaffilmark{1,2}, 
Yasushi Suto\altaffilmark{1}}

\affil{Department of Physics, The University of Tokyo, Tokyo 113-0033,
Japan}

\email{ohta@utap.phys.s.u-tokyo.ac.jp,
ataruya@utap.phys.s.u-tokyo.ac.jp, suto@phys.s.u-tokyo.ac.jp}

\altaffiltext{1}{also at Research Center for the Early Universe 
(RESCEU), School of Science, The University of Tokyo, Tokyo 113-0033,
Japan.}
\altaffiltext{2}{
also at Institute for the Physics and Mathematics of the Universe,
University of Tokyo, Kashiwa, Chiba 277-8568, Japan}

\begin{abstract}
We present theoretical predictions for photometric and spectroscopic
signatures of rings around transiting extrasolar planets.  On the basis
of a general formulation for the transiting signature in the stellar
light curve and the velocity anomaly due to the Rossiter effect, we
compute the expected signals analytically for a face-on ring system, and
numerically for more general configurations.  We study the detectability
of a ring around a transiting planet located at $a=3$AU for a variety of
obliquity and azimuthal angles, and find that it is possible to detect
the ring signature both photometrically and spectroscopically unless the
ring is almost edge-on (i.e., the obliquity angle of the ring $\theta$
is much less than unity).  We also consider the detectability of
planetary rings around a close-in planet, HD 209458b ($\theta\approx
90^\circ-i_{\rm orbit} \approx 3^\circ.32$), and Saturn ($\theta \approx
26^\circ.7$) as illustrative examples. While the former is difficult to
detect with the current precision (photometric precision of $10^{-4}$
and radial velocity precision of 1 m s$^{-1}$), a marginal detection of
the latter is possible photometrically. If the future precision of the
radial velocity measurement reaches even below 0.1 m s$^{-1}$, they will
be even detectable from the ground-based spectroscopic observations.
\end{abstract}
\keywords{planets:, techniques:spectroscopic}

\section{Introduction}

In 1903, Hantaro Nagaoka, a professor in physics, Imperial University of
Tokyo, proposed an atomic model which {\it consists of a large number of
particles of equal mass arranged in a circle at equal angular intervals}
around a central particle of large mass \citep{nagaoka03,nagaoka04}. He
referred to this model as a {\it Saturnian} model, and published the
hypothesis prior to the famous Rutherford model \citep{rutherford11},
often referred to as a {\it planetary} model of the atom. This remarkable
example in history illustrates how our Solar system, and even a
Saturnian ring, inspired the best physicists in the world to contemplate
the quantum world.

Thus it is interesting to contemplate how the properties of more than
300 extrasolar planets discovered so far beyond our Solar system will
inspire future scientists.  Among them, 52 systems are known to exhibit
transiting signatures. Detailed photometric and spectroscopic
observations of such transiting planets provide unique opportunities to
look into the nature of extrasolar planets, which is otherwise
unaccessible including the planetary radius, average density, and
atmospheric composition.  Ongoing/upcoming missions such as COROT and
Kepler will significantly increase the number of transiting planets, and
even a wider range of research will be made possible in the next several
years. One specific example is the exploration of the spin-orbit
misalignment of extrasolar planet using the Rossiter effect
\citep{Rossiter24, McLaughlin24, Petrie38, Kopal42, Kopal45,
Hosokawa53}; when a planet transits its host star, the light coming from
the star is partially blocked.  Photometrically this produces a
characteristic dimming of the stellar light curve, and
spectroscopically, an asymmetric distortion of the absorption line
profiles of the stellar spectrum due to the occultation of some of the
velocity components of the stellar absorption lines.  Thus the latter
yields an {\it apparent} anomaly in the radial velocity curve of the
host star \citep{Queloz00}.

In a previous paper \citep[][hereafter Paper I]{OTS05}, we derived
analytic templates for the velocity anomaly using a perturbative
expansion. Motivated by that work, \citet{Winn05} combined the best
observational datasets currently available for the HD 209458 system, and
detected for the first time a small misalignment ($\lambda= -4^\circ.4
\pm 1^\circ.4$) between the spin axis of the central star and the
orbital axis of the planet. Subsequently \citet{Wolf06} applied the
analytic formulae of Paper I, and obtained the limit, $\lambda=
+11^\circ \pm 14^\circ$, for the anomalously dense planet, HD 149026
\citep{bsato05}.  Also, the spin-orbit alignment of the HD 189733 system
were measured by \citet{Winn06} and a very small misalignment of
$\lambda=-1^{\circ}.4\pm1^{\circ}.1$ was found \cite[][see
also]{Winn07a,Winn07b,Narita07}.  Recently measurements of spin-orbit
alignment have been successful for a number of transiting planets.  It
is just a matter of time to obtain the statistical distribution of the
misalignment angle $\lambda$, which provides a unique probe of the
origin and evolution of the angular momentum of the planetary systems.

In the present paper, we consider an even more ambitious possibility to
detect rings around transiting planets.  It is interesting to note that
the rings around Uranus and Neptune were indeed discovered during their
occultation
of background stars
\citep{Elliot77,Hubbard77,Hubbard86,esposito06}.  While it is not clear
whether or not extrasolar planets have rings and/or satellites as those
in our solar system, their detection, if successful, would definitely
mark an important milestone in astronomy and planetary sciences;
transiting planets are the unique targets for that purpose.  From the
methodology developed in this paper, one can further determine the
orientation of the spin of the orbiting planet under a reasonable
assumption that the orientation of the ring is aligned to the planetary
spin axis; this is approximately the case for planetary rings in our
Solar system.

Indeed we are not the first to explore the detectability in detail.
\citet{Brown01}, for instance, put photometric constraints on possible
satellites and rings around the first transiting planet, HD 209458b.
They assume that an optically thick two-dimensional ring extending from
the surface of the planet up to the outer radius $R_{\rm out}$.  Hot
Jupiters like HD 209458b are supposed to be roughly tidally locked to the
host star. Thus they also assume that the ring plane coincides with the
planetary orbital plane, and found that $R_{\rm out}$ should be less
than 1.8 times the planetary radius. Given the approximate edge-on
nature of the assumed ring, the upper limit is fairly strong and
interesting. More recently \citet[][hereafter BF04]{BF04} presented a
light curve of ringed planets over a wide range of the parameter space,
and discussed the photometric detectability of the additional ring
signature.

Our present paper differs from the previous work in that we examine the
{\it spectroscopic} detectability of planetary rings using the Rossiter
effect in addition to the photometric one.  Given the current
observational sensitivities, the photometric signatures are easier to
detect than the spectroscopic ones in most cases. Nevertheless, the
complementary nature of the spectroscopic detection/confirmation of
rings is important because the signature of rings is very subtle and may
be mimicked/disguised by unknown noise and systematics. Moreover the
spectroscopic signature may be more pronounced for transiting planets in
rapidly rotating stars because the velocity anomaly due to the Rossiter
effect is simply proportional to the stellar spin velocity.

We also present improved and more accurate analytic formulae of the
photometric light curve and the spectroscopic velocity anomaly for a
transiting planet {\it without} a ring than those in Paper I.  While
those for a planet with a ring need to be computed numerically in
general, such analytic formulae are very useful in performing the
parameter fit for a ringless planet model, and also in checking the
accuracy of the numerical results for a planet with a ring.

The plan of the paper is as follows; we describe our basic assumptions
for the planet and ring system in \S 2.  Section 3 summarizes the
general formulation to compute the photometric and spectroscopic signals
for the planet and ring system. Analytical results for the face-on ring
are presented in \S 4, which are in fact the improved version of the
results of Paper I. Numerical results for an arbitrarily oriented ring
are shown in \S 5, and we discuss its detectability in detail.  Finally
\S 6 is devoted to conclusions and implications of the paper.

\section{Model assumptions}

Let us begin with describing several definitions and assumptions
concerning the geometrical configuration of the planetary system and the
properties of the host star, the orbiting planet and a surrounding
ring.  Table \ref{tbl:parameters} lists the notations and the parameters
of our model, which are used in the analysis below.

\begin{figure}[hbt]
\epsscale{0.4}
\plotone{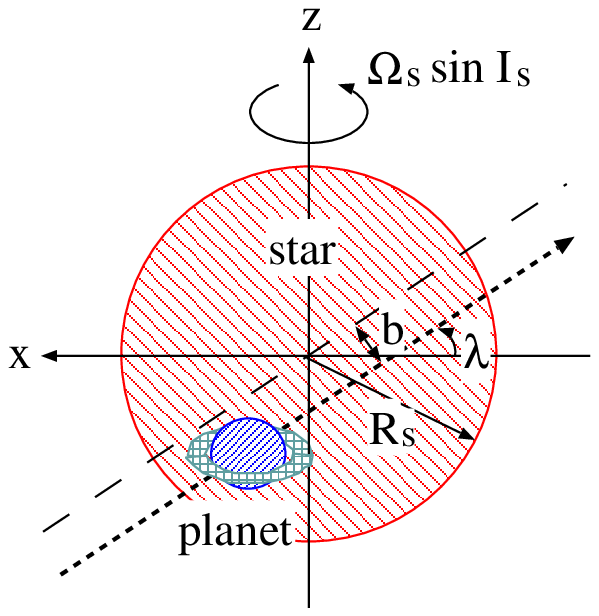}
\figcaption{Planetary transit accompanied by a ring.  The internal
parameters of the star and the planet are shown: stellar radius $R_s$;
projected angular velocity $\Omega_s\,\sin I_s$; impact parameter of
planetary orbit, $b$; angle between the $x$-axis and the orbital plane,
$\lambda$, which represents the spin-orbit alignment projected onto the
$x-z$ plane.  \label{fig:orbit}}
\end{figure}

\subsection{Star--planet system}

We assume that the star (radius $R_s$) and the planet (radius $R_p$)
follow the exact two-body Kepler orbit, ignoring perturbations from
possible outer planets.  We adopt a coordinate system centered at the
star so that its $y$-axis is directed toward the observer, and the
$z$-axis is chosen so that the stellar rotation axis lies on the $y$-$z$ 
plane.  We assume rigid-body stellar rotation with $\Omega_s$ being its
spin angular velocity.  In this coordinate system, the components of the
stellar spin angular velocity reduces to
\begin{equation}
   {\bf \Omega}_s = (0, \Omega_s \cos I_s, \Omega_s \sin I_s),
\end{equation}
where $I_s$ is the inclination angle between the stellar spin axis and
observer's direction.

 Since we are mainly interested in the transit phase whose duration is
much shorter than the entire orbital period, the actual eccentric orbit
is close to a circular one in the vicinity of the planetary transit.
Thus we adopt the circular orbit in the present analysis while most of
our formulae below are applicable to a non-circular orbit by properly
mapping the projected position of the planet into the observed time.
Under the circular orbit approximation, the planetary orbit is simply
characterized by the impact parameter $b$ and the projected angle
$\lambda$ between the stellar spin and the planetary orbital axes, in
addition to the semi-major axis $a$ (Fig. \ref{fig:orbit}).  The impact
parameter $b$ is equal to $b=a \cos i_{\rm orbit}/R_s$, where $i_{\rm
orbit}$ is the inclination of the orbit.

We adopt the linear limb darkening alone for the surface intensity of a
stellar disk in Paper I. In this paper, we take account of the effect up
to quadratic order:
\begin{equation}
   I_{\rm star}(\mu)=I_0\left[1-u_1(1-\mu)-u_2(1-\mu)^2\right].   
   \label{eq:limbdarkening}
\end{equation}
The two constants, $u_1$ and $u_2$, characterize the amplitude of the
limb darkening, and $\mu$ is the directional cosine between the line of
sight and the normal vector to the local stellar surface:
\begin{equation}
\mu= \sqrt{1-\frac{x^2+z^2}{R_s^2}}.  
\label{eq:def_of_mu}
\end{equation}

\subsection{Planet and ring \label{subsec:planet_ring}}

\begin{figure}[htb]
\epsscale{0.4}
\plotone{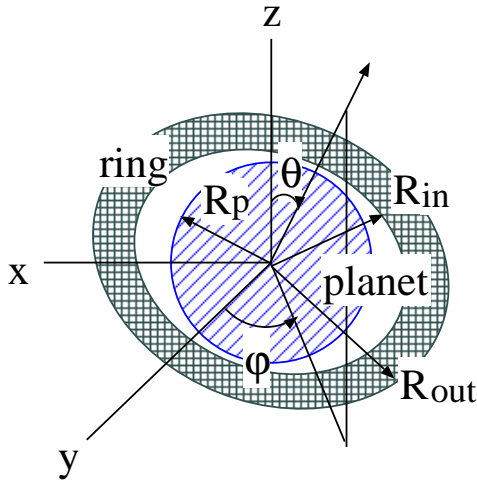}
\figcaption{ Schematic configuration of the planet accompanied by a ring.
The parameters of the planet and the ring include planetary radius
$R_p$, radius of inner edge of ring $R_{\rm in}$, radius of outer edge
of ring $R_{\rm out}$, obliquity angle $\theta$ (the angle
between the normal vector to the ring surface and $z$-axis), rotation
angle $\phi$ (the angle between the $y$-axis and the projected
normal vector of ring surface onto the $y-z$ plane).
\label{fig:ring-direction}}
\end{figure}

We consider a simple model of ring around a planet as shown in Figure
\ref{fig:ring-direction}. The ring is two-dimensional (geometrically
thin) and circular.  Its inner and outer radii are denoted by $R_{\rm
in}$ and $R_{\rm out}$ which satisfy $R_p<R_{\rm in}<R_{\rm out}$. We
assume that the ring moves along the planetary orbit with constant
obliquity angles $(\theta,\phi)$.  The light transmitted through the
ring diminishes due to the absorption and/or the extinction by particles
in the ring. We introduce the normal optical depth of ring, $\tau$, so
that the transmitted flux through the ring is proportional to
$\exp(-\tau/\beta)$, where $\beta$ is the cosine of the angle between
the normal vector to the ring surface and $y$-axis, and given by
$\beta=\sin\theta\cos\phi$. We adopt $\tau=1$
for simplicity (see Table \ref{tbl:parameters}); the optical depth of
the B ring of Saturn is in the range of $\tau=0.4 \sim 2.5$ \citep{SSD}.

\cite{BF04} examined the influence of diffraction on the photometric
light curves, and found that it may leave a detectable signature
depending on the size of ring particles.  We ignore the diffraction of
the ring particles here for two reasons; first, our primary purpose is
to develop a methodology which directly tests the presence of rings, and
second, the diffraction effect itself does not exceed the primary effect
of the star-light dimming. In any case the effect of diffraction can be
easily incorporated into our numerical scheme if the ring properties are
specified.

\section{Photometric and spectroscopic
 transit signatures for a planet with a ring}

\subsection{Photometric signals}

  The photometric light curve for transiting extrasolar planet with a
ring has been computed by \citet{BF04} in detail. For definiteness and
later convenience, we summarize their basic equations in terms of our own
notations here.

The dimming of the star light during the transit phase is quantified by
evaluating the fraction of star light blocked by the planet and/or
intercepted the ring. We define the relative flux ratio $F$:
\begin{equation}
   \label{eq:flux_ratio}
   F =\frac{\displaystyle \iint I(x,z) dxdz}
   {\displaystyle \iint I_{\rm s}(x,z) dxdz}.  
\end{equation}
Here, the functions $I(x,z)$ and $I_{\rm s}(x,z)$ represent an intensity
at a given position $(x,z)$ of the stellar surface.  The stellar
intensity model (\ref{eq:limbdarkening}) and the ring model described in
\S \ref{subsec:planet_ring} yield the explicit form of the function
$I(x,z)$:
\begin{eqnarray}
I(x,z)=\left\{
       \begin{array}{lcl}
    0 &:& \mbox{region outside the star, i.e.,~ }x^2+z^2>R_s^2  \\
    0 &:& \mbox{region inside  the planet, i.e.,~ }
    (x-X_p)^2+(z-Z_p)^2\leq R_p^2 \\
    e^{-\tau/\beta}I_{\rm star}(\mu) &:& 
    \mbox{region transmitted through the ring, see} 
~~{\rm conditions\,\,(\ref{eq:ring_transmit_cond})}\\
    I_{\rm star}(\mu) &:& \mbox{otherwise}
\end{array}
\right.,
\label{eq:intensity_I}
\end{eqnarray}
where the direction cosine $\mu$ is given by equation (\ref{eq:def_of_mu}). 
The region where the light coming from the star is transmitted through
the ring satisfies the three conditions:
\begin{eqnarray}
\left\{
       \begin{array}{l}
 R_{\rm in}^2<\left(\displaystyle 
\frac{x-X_p-(z-Z_p)\sin\phi}{\cos\phi}\right)^2
    +\left(\displaystyle 
\frac{z-Z_p}{\sin\theta}\right)^2<R_{\rm out}^2, \\
    x^2+z^2<R_s^2, \\
 (x-X_p)^2+(z-Z_p)^2>R_p^2 
\end{array}
\right.
\label{eq:ring_transmit_cond}
\end{eqnarray}
The quantities, $X_p$ and $Z_p$, represent the position centered at the
planet, which manifestly depend on time.

The stellar intensity outside the transit phase, $I_{\rm s}(x,z)$, is
simply written as
\begin{eqnarray}
   I_{\rm s}(x,z)=
   \left\{
       \begin{array}{lcl}
    0 &:& \mbox{outside the star i.e., }x^2+z^2>R_s^2\\
    I_{\rm star}(\mu) &:& \mbox{inside the star i.e., }x^2+z^2<R_s^2
    \end{array}
    \right..
\end{eqnarray}
Substituting the stellar intensity model (\ref{eq:limbdarkening}), the
denominator of equation (\ref{eq:flux_ratio}) can be integrated
analytically:
  \begin{equation}
   \iint I_s(x,z) dxdz = \pi 
   I_0R_s^2\left(1-\frac{u_1}{3}-\frac{u_2}{6}\right).
  \end{equation}
Note that this is independent of time.

\subsection{Spectroscopic signals}

While planetary rings may be detectable with photometric data analysis
alone, the spectroscopic data provide an independent and complementary
check of the presence of rings. Since the two methods would suffer from
very different noises and systematics, the agreement between them 
significantly increases the credibility of the positive detection.

The spectroscopic detection method relies on the Rossiter effect
\citep{Rossiter24} which represents the ``apparent'' radial velocity
modulation arising from the spectral line distortion due to an
occultation of a part of the rotating star. During the passage of the
transiting planet, the light from the stellar surface is asymmetrically
blocked off. As a result, the line-profile-weighted mean position for a
specific absorption or emission line taking account of the stellar
rotation is apparently shifted.  In Paper I, we derived analytical
expressions for radial velocity curves for extrasolar transiting planets
for the first time, but the effect of rings was ignored. The
quantitative modelling of the photometric and spectroscopic signatures
of rings will be considered in detail below.

Formally the radial velocity anomaly due to the Rossiter effect
is expressed as
  \begin{equation}
   \Delta v_s = -\Omega_s \sin I_s
   \frac{\displaystyle \iint x\,\,I(x,z) dxdz}
   {\displaystyle \iint I(x,z) dxdz}.
   \label{eq:rad_vel_shift}
  \end{equation}
The above expression, which relates the velocity anomaly to the first
moment of the line profile, is applicable also to transiting planets
with a ring if equation (\ref{eq:intensity_I}) is substituted into
$I(x,z)$.  In general, the above integral cannot be performed
analytically but needs to be done numerically (except for a face-on ring
as described in \S \ref{sec:face-on}).  We note here that \citet{Winn05}
pointed out the presence of $\sim 10$ percent systematic difference
between the perturbative results of Paper I and their numerical fits to
the outputs of the data analysis pipeline. This is supposed to come from
the fact that the approximation of the first moment of the line profile
as $\Delta v_s$ is not sufficiently accurate. The systematic difference
should be sensitive to the specific data analysis routine, and hard to
evaluate generally. Moreover it is unlikely to affect our conclusions
which depend on the {\it difference} of $\Delta v_s$ of planets with and
without rings. This is why we use equation (\ref{eq:rad_vel_shift}) 
to evaluate the Rossiter effect in what follows.

In practice, we proceed as follows; we first calculate the integral over
an entire surface of the stellar disk, which can be done analytically.
Then we numerically evaluate the integrals over the regions overlapped
with the planet and the ring, and subtract these from the analytic
result without a planet nor a ring. In order to cope with the
complicated geometry of the integration, we set square grids ($256\times
256$ cells) enclosing both the planet and the ring with a length of
$2R_{\rm out}$ on a side. Judging from the location of the center of
each cell relative to the stellar surface, we assign the intensity
$I(x,z)$ in equation (\ref{eq:intensity_I}), and compute the
contribution from the cell.  We test the convergence and the accuracy of
our numerical integrations by comparing with the results with finer
grids ($512\times 512$ cells). We find that the fractional difference of
the photometric signal is less than $3\times 10^{-5}$, and that the
spectroscopic signals agree within $0.08{\rm ms^{-1}}$, both of which are
negligibly small for our current purpose.

\section{Analytic formulae for a planet with a face-on ring 
\label{sec:face-on}}

Equations (\ref{eq:flux_ratio}) and (\ref{eq:rad_vel_shift}) can be
integrated (almost) analytically for a planet with a face-on ring, i.e.,
$(\theta,\,\phi)=(90^{\circ},\,0^{\circ})$.  These analytic formulae
help our understanding of the basic observational signatures of a ring
from the photometric and the spectroscopic data. They are also useful in
making sure of the accuracy of the numerical results for more general
cases.

We find that equation (\ref{eq:flux_ratio}) reduces to
\begin{equation}
   F  = 
   \frac{\displaystyle \pi(1-u_1/3-u_2/6)-A(\rho,\gamps)-(1-e^{-\tau})
     \left\{A(\rho,\gamo)-A(\rho,\gami)\right\}}
   {\pi(1-u_1/3-u_2/6)}
   \label{eq:face-on_1}
  \end{equation}
for the relative flux ratio, and equation (\ref{eq:rad_vel_shift}) to 
  \begin{equation}
   \Delta v_s=X_p\,\,\Omega_s\,\sin I_s\,\,
   \frac{\displaystyle B(\rho,\gamps)+(1-e^{-\tau})
     \left\{B(\rho,\gamo)-B(\rho,\gami)\right\}}
   {\displaystyle \pi(1-u_1/3-u_2/6)-A(\rho,\gamps)-(1-e^{-\tau})
     \left\{A(\rho,\gamo)-A(\rho,\gami)\right\}}. 
   \label{eq:face-on_2}
  \end{equation}
for the radial velocity shift, where $\gamps=R_p/R_s$, $\gami=R_{\rm
in}/R_s$, $\gamo=R_{\rm out}/R_s$, and
$\rho\equiv\sqrt{X_p^2+Z_p^2}/R_s$.  The quantities, $A$ and $B$, are 
given in terms of the normalized planet position $\rho$ as follows:
\begin{eqnarray}
\label{eq:func-A}
A(\rho,\,\gamma)=
   \left\{
   \begin{array}{lcl}
    \pi\gamma^2
    \left[1-u_1-u_2\left(2-\rho^2-\gamma^2/2\right)+
        (u_1+2u_2)W_1\right] &:& \rho<1-\gamma 
\\
\\
    \left(1-u_1-3u_2/2\right)\left[\gamma^2\cos^{-1}
        \left(\zeta/\gamma\right)+\sin^{-1}z_0-\rho z_0\right]&&
\\
    \quad+(u_2/2)\,\rho\left[(\rho+2\zeta)\gamma^2
        \cos^{-1}\left(\zeta/\gamma\right)-z_0
        \left(\rho\zeta+2\gamma^2\right)\right]
    &:& 1-\gamma\leq\rho\leq 1+\gamma 
\\
    \quad\quad+(u_1+2u_2)W_3 &&
\\
\\
   0 &:& 1+\gamma<\rho
    \end{array}
    \right.
\end{eqnarray}
and 
\begin{eqnarray}
   B(\rho,\,\gamma)=
   \left\{
     \begin{array}{lcl}
    \pi\gamma^2\left[1-u_1-u_2\left(2-\rho^2-\gamma^2\right)+
        (u_1+2u_2)W_2\right] &:& \rho<1-\gamma 
\\
\\
    \left[1-u_1-u_2\left(2-\rho^2-\gamma^2\right)\right]
    \left[\gamma^2\cos^{-1}\left(\zeta/\gamma\right)-z_0\zeta\right]
     && 
\\
\quad -(4u_2/3)\rho z_0^3+(u_1+2u_2)W_4/\rho
     &:& 1-\gamma\leq \rho\leq 1+\gamma 
\\
\\
    0 &:& 1+\gamma<\rho
     \end{array}
     \right.
\end{eqnarray}
The variables $z_0$ and $\zeta$ are also written in terms of 
$\gamma$ and $\eta_p=\rho-1$ as
  \begin{equation}
   z_0(\gamma)=\frac{\sqrt{(\gamma^2-\eta_p^2)
       \Bigl[(\eta_p+2)^2-\gamma^2\Bigr]}}{2(1+\eta_p)},
\quad\quad
   \zeta(\gamma)=\frac{\eta_p^2+2\eta_p+\gamma^2}{2(1+\eta_p)}.
  \end{equation}
The functions $W_i$ ($i=1 \sim 4$) in the above expressions are in fact
analytically intractable, but we find accurate analytical fitting
formulae which are explicitly given in Appendix \ref{appendix:formula}.

If we set $\tau=0$, the above analytic expressions reduce to a case of a
planet without a ring.  However they are superior to, and thus should
replace, our previous results \citep{OTS05} in two ways; first, we now
consider a quadratic limb darkening (eq.[\ref{eq:limbdarkening}]),
instead of a linear limb darkening ($u_2=0$), and second, the
approximate expressions for $W_i$ given in the appendix are more
accurate by a factor of a few than the previous one; for typical
parameters $\gamps\leq\gami\leq\gamo\sim0.2$, the accuracy is typically
within a few percent level.  We use the above formulae extensively when
we discuss the effect of rings in the following analysis.

\section{Predicted photometric and spectroscopic signals}

Even in the simple model that we assume here, the photometric light
curve and the spectroscopic velocity curve are characterized by 12 and
14 parameters, respectively.  The former includes the radii of the star
and the planet ($R_s$ and $R_p$), the inner and the outer edges of the
ring ($R_{\rm in}$ and $R_{\rm out}$), the limb-darkening parameters
($u_1$ and $u_2$), the orbital parameters of the system (semi-major axis
$a$, inclination of the orbital plane $i_{\rm orbit}$, and the orbital
rotation period $P_{\rm orbit}$), the obliquity angles of the ring
($\theta$ and $\phi$), and finally the normal optical depth of the ring,
$\tau$. The additional two parameters for the latter are the projected
stellar surface velocity $V\sin I_s$ and the projected misalignment
angle $\lambda$ between the stellar spin and the orbital plane axes.

In order to avoid the tidal disruption, the outer radius of the ring
should not exceed the Hill radius:
\begin{equation}
R_{\rm out} < r_{\rm H} = \left(\frac{M_p}{3M_s}\right)^{1/3} a
\approx 15.4 R_J \left(\frac{M_p}{M_J}\right)^{1/3}
\left(\frac{M_\odot}{M_s}\right)^{1/3}
\left(\frac{a}{0.05{\rm AU}}\right).
\end{equation}
\citet{Gaudi03} discussed further theoretical constraints on planetary
rings. Planetary rings consisting of ice sublimation if the planetary
semi-major axis satisfies
\begin{equation}
 a > \left(\frac{L_*}{16\pi\sigma T_{\rm sub}^4}\right)^{1/2}
=2.7{\rm AU}\left(\frac{L_*}{L_\odot}\right)^{1/2} ,
\end{equation}
where $T_{\rm sub}=170{\rm K}$ is the sublimation temperature of water
ice, and $L_*$ is the luminosity of the host star. For rocky rings, the
Poynting Robertson drag force, viscous friction due to the planetary
exosphere, torques due to (shepherd) satellites, and internal scattering
become important. For close-in planets, the Poynting Robertson drag
time-scale $t_{\rm PR}$ is generally shorter than the viscous friction
decay time-scale unless the density of the exosphere is large,
$\rho_{\rm es}\simgt 10^{-16}{\rm g cm^{-3}}$ \citep[][]{GT82}. The former is
given as
\begin{equation}
 t_{\rm PR}\sim 10^5 {\rm yr}\frac{\rho}{1{\rm g cm^{-3}}}
\frac{r}{1{\rm cm}}\left(\frac{a}{0.1{\rm AU}}\right)^2,
\end{equation}
where $\rho$ and $r$ denote the density and the radius of ring
particles. Thus rings around close-in planets would not live longer than
$t_{\rm PR}$ unless any stabilization mechanism like shepherd satellites
operates.  Thus it is not clear if close-in planets are accompanied by
rings.

Based on these estimates, we adopt a fiducial set of parameters listed
in Table \ref{tbl:parameters}, which is motivated by the existing
close-in planetary system HD 209458, but with different semi-major axis,
$a=3$AU.  As illustrative examples, we also consider the close-in planet
and the Saturn system as a member of solar-system planet.  We first
present a special case of the face-on ring (that can be analytically
computed) so as to elucidate the basic features (\S
\ref{subsec:faceon-ring}), and to quantify the parameter dependence of
the photometric and spectroscopic signals in \S \ref{subsec:parameters}.
Next in \S \ref{subsec:signature}, we discuss how to identify the
signature of rings by looking for the departure from the fit to a planet
model without ring.  Based on these discussions, the detectability of a
ring system is addressed in \S \ref{subsec:detectability}, in addition
to future prospects for constraining the ring parameters.

\subsection{Signature of the face-on ring \label{subsec:faceon-ring}}

\begin{figure}[htb]
    \epsscale{0.5} \plotone{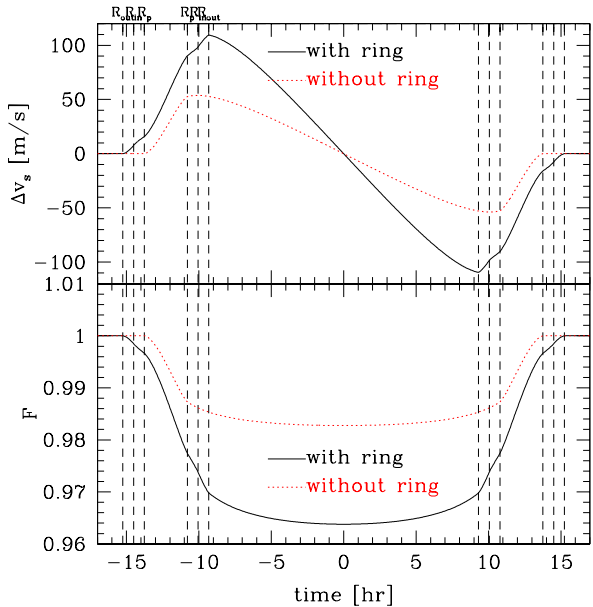} 
\figcaption{ Schematic illustration for photometric and 
  spectroscopic signatures of transiting planet with and without 
  a ring (see Table \ref{tbl:parameters} 
  for the fiducial values of the model parameters). 
  The {\it top panel} shows the velocity anomaly curve for
  the Rossiter effect. The ring is assumed to be
  face-on, i.e.,  $\theta=90^\circ$, $\phi=0^\circ$. The curve is 
  analytically calculated according to the expression
  (\ref{eq:face-on_2}).  The {\it bottom panel} shows the light curve
  during transit. The ring is  also face-on. The curve is 
  analytically calculated according to the expression 
  (\ref{eq:face-on_1}).  \label{fig:faceon}}
\end{figure}

Figure \ref{fig:faceon} shows the velocity anomaly by the Rossiter
effect ({\it upper}) and the light curve of photometry ({\it lower}) for
a planet with a face-on ring, i.e., $\theta=\varphi=0$.  The central
transit time is chosen as $t=0$.  These curves are plotted using the
analytical expressions (\ref{eq:face-on_1}) and (\ref{eq:face-on_2}).

Compared to the curves of a planet without a ring ({\it dashed lines}),
the planetary ring produces larger radial velocity shifts and flux
dimming ({\it solid lines}).  Also the transit duration increases.  As
we will show below, however, most of those features due to the ring
resemble that of a ringless planet with a larger radius.  The vertical
dotted lines in Figure \ref{fig:faceon} indicate the epochs when the
boundary edges of the planet and the ring contact the edge of the
stellar disk (i.e., the ingress and egress phases). These somewhat
discontinuous features in the derivatives of the curves, if detectable
directly at all, would be an unambiguous confirmation of the presence of
the ring. In reality, however, it may be difficult to locate the
features from the sampled data with a finite exposure time.

A more realistic methodology is to first search for the best-fit
parameters assuming no ring at all, and then to look for any systematic
ring signatures in the residuals between the observed data and the fits
\citep{BF04}. Even so, the detection of rings requires better
resolutions in sampling time, as well as in photometric and spectroscopic
accuracy, than those of current data.

\subsection{Dependence on the ring parameters \label{subsec:parameters}}

\begin{figure}[htb]
    \epsscale{0.5} \plotone{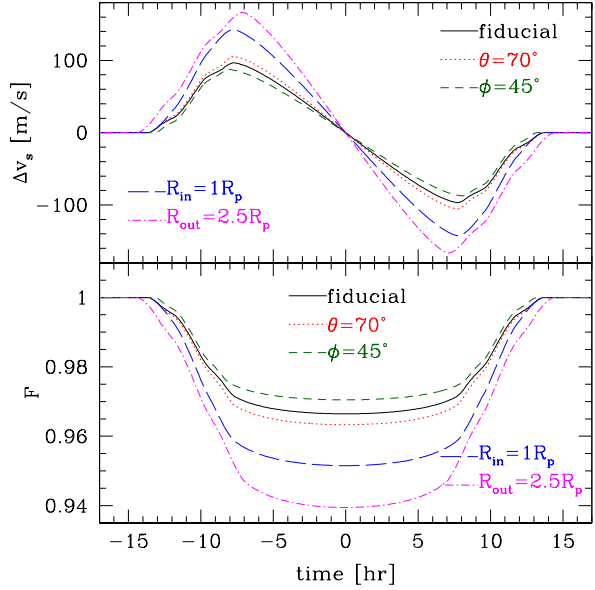} \figcaption{ Same as Figure
  \ref{fig:faceon}, but for a ring with different set of parameters.
  The top panel shows the velocity anomaly from Rossiter effect, while
  bottom panel plots the light curve during transit. In each panel,
  solid lines represent the results for the fiducial model (Table
  \ref{tbl:parameters}).  Other lines show the results when one of the
  fiducial parameters is varied; $\theta=70^\circ$ (dotted),
  $\phi=45^\circ$ (short-dashed), $R_{\rm in}=R_p$ ({\it long-dashed}),
  and $R_{\rm out}=2.5R_p$ ({\it dot-dashed}).  \label{fig:parameter
  dependence}}
\end{figure}
\begin{figure}[htb]
    \epsscale{0.5}
  \plotone{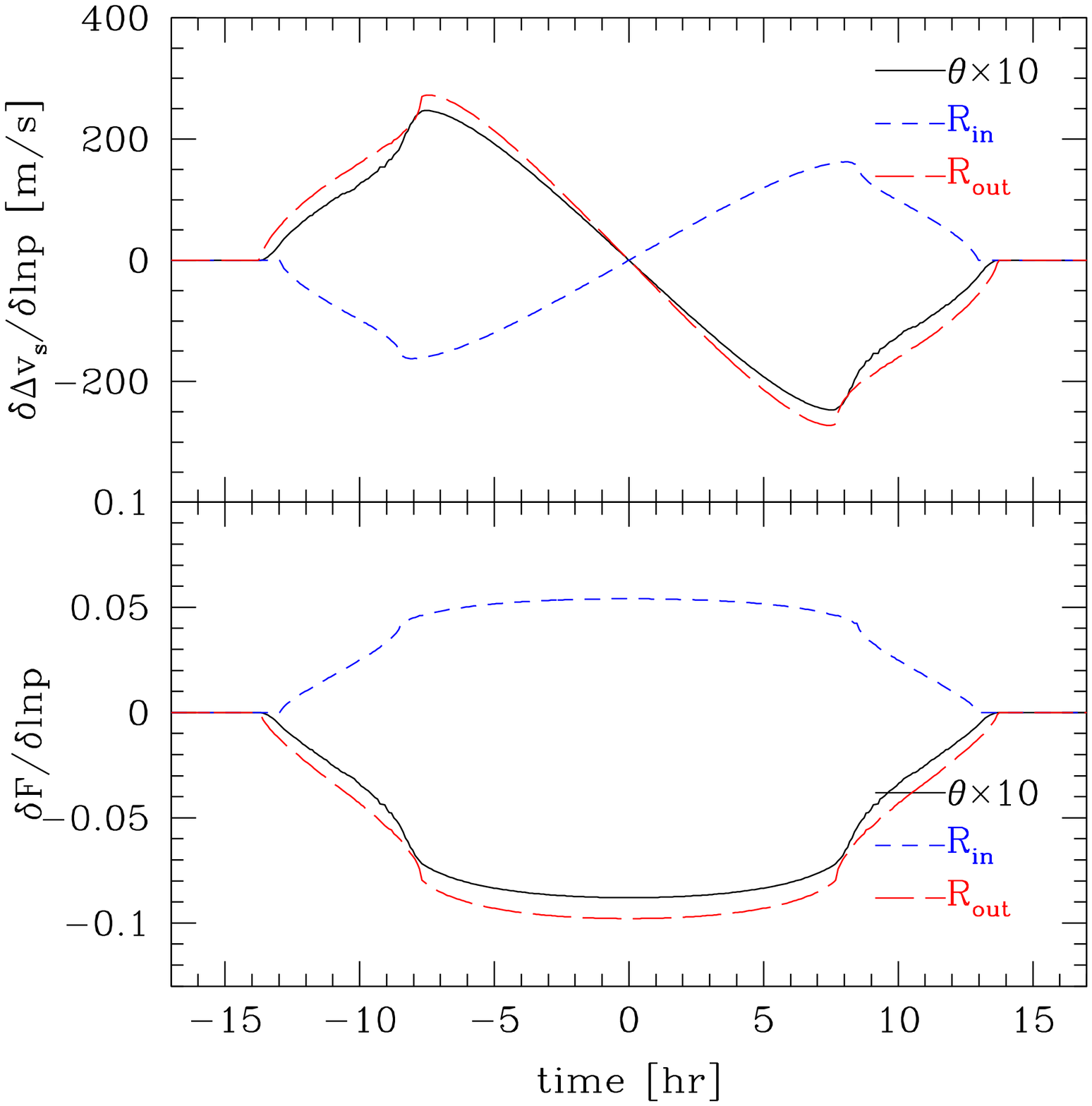}
  \figcaption{Variation of the photometric ({\it upper}) and spectroscopic
 ({\it lower}) signals with respect to the ring parameters, $R_{\rm in}$, 
 $R_{\rm out}$, and $\theta$. The curves for the latter case are 
 multiplied by a factor of 10 for clarity.
  \label{fig:parameter differential}}
\end{figure}

In general, the photometric and the spectroscopic signals due to rings
need to be evaluated numerically. Here we discuss their sensitivity on
the four ring parameters ($\theta, \varphi, R_{\rm in}$ and $R_{\rm
out}$).  Figure \ref{fig:parameter dependence} shows some examples with
slightly different parameters from our fiducial values; $\theta=45^\circ
\to 70^{\circ}$ ({\it dotted}), $\phi=0^\circ \to 45^{\circ}$ ({\it
short-dashed}), $R_{\rm in}=1.5R_p \to 1.0R_p$ ({\it long-dashed}) and
$R_{\rm out}=2.0R_p \to 2.5R_p$ ({\it dot-dashed}).

In order to clarify how the predicted $\Delta v_s$ and $F$ are sensitive
to the underlying parameters of the system ($\theta$, $\phi$, $R_{\rm
in}$ and $R_{\rm out}$), we compute their dependence as we did in Figure
9 of Paper I. Specifically, Figure \ref{fig:parameter differential}
illustrates the variation of the radial velocity shifts $\Delta v_s$
with respect to $p=R_{\rm in}$, $R_{\rm out}$ and $\theta$:
\begin{equation}
   \frac{\delta\Delta v_s}{\delta \ln p} \equiv \lim_{dp\rightarrow 0}
    \frac{\Delta v_s(p+dp;t)-\Delta v_s(p;t)}{dp/p} .
\end{equation}
Similarly we define the variation of the photometric light curves
$\delta F$:
\begin{equation}
   \frac{\delta F}{\delta \ln p} \equiv \lim_{dp\rightarrow 0}
    \frac{F(p+dp;t)-F(p;t)}{dp/p} ,
\end{equation}
which are also plotted in Figure \ref{fig:parameter
differential}.  Since the dependence on $\theta$ is small, we
multiply the values of $\delta \Delta v_s/\delta \ln \theta$ and 
$\delta F/\delta\ln\theta$ by a factor of 10 in the plot.

Figure \ref{fig:parameter dependence} indicates that a different set of
the ring parameters mainly changes the amplitude of signals during the
full transiting phase, and that the bumpy structures around the ingress
and the egress phases, which indeed are the important signatures of
possible rings, are not affected much. This can be understood from the
weak dependence on the angle parameters in Figure \ref{fig:parameter
differential}. If we changes the value of one of the ring parameters
from $p$ to $p'$, the photometric and spectroscopic signals become
\begin{eqnarray}
F(p';t) &\approx& 
F(p;t) + \left(\frac{p'-p}{p}\right)\frac{\delta F}{\delta \ln p}, \\
\Delta v_s(p';t) &\approx& 
\Delta v_s(p;t) + \left(\frac{p'-p}{p}\right)
\frac{\delta \Delta v_s}{\delta \ln p} .
\end{eqnarray}
Those approximate perturbation formulae combined with Figure
\ref{fig:parameter differential} explain the behavior of Figure
\ref{fig:parameter dependence} fairly well, in particular the weak
dependence on $\theta$.

We note that the three ring parameters ($\theta, R_{\rm in}$ and $R_{\rm
out}$) symmetrically change the photometric light curves around the
central transit time and anti-symmetrically change the radial velocity
shifts (Fig.\ref{fig:parameter differential}). Such behavior is quite
similar to the variation with respect to the planetary and the stellar
internal parameters (see Fig.9 of Paper I). Thus, the overall change of
the amplitude is not a sensitive measure of the presence of a
ring. Therefore the precise measurement of the bumps around the ingress
and the egress phases, even if difficult, provides the key in the
detection and the characterization of the planetary ring.

\subsection{Signatures of rings \label{subsec:signature}}

The results of the previous subsections now enable us to discuss in
detail the signatures of a ring system in photometric and
spectroscopic data.
Given real data set, first thing that we should do is
to fit the photometric light curve and/or
the spectroscopic velocity anomaly of a transit planet using a model
without ring. If the residuals between the data and the best-fit model
are significantly larger than the observational
errors, one suspects the presence of a ring.

More specifically, let us introduce the above residuals as
\begin{equation}
    \mbox{Res}[v_s]\equiv
    \Delta v_s^{\rm obs}-\Delta v_s^{\rm best-fit}~(\mbox{without ring})
    \label{eq:Res_v_s}
\end{equation}
and 
\begin{equation}
    \mbox{Res}[F]\equiv
    F^{\rm obs}-F^{\rm best-fit}~(\mbox{without ring}) .
    \label{eq:Res_F}
\end{equation}

\begin{figure}[htp]
 \epsscale{1.1}
 \plottwo{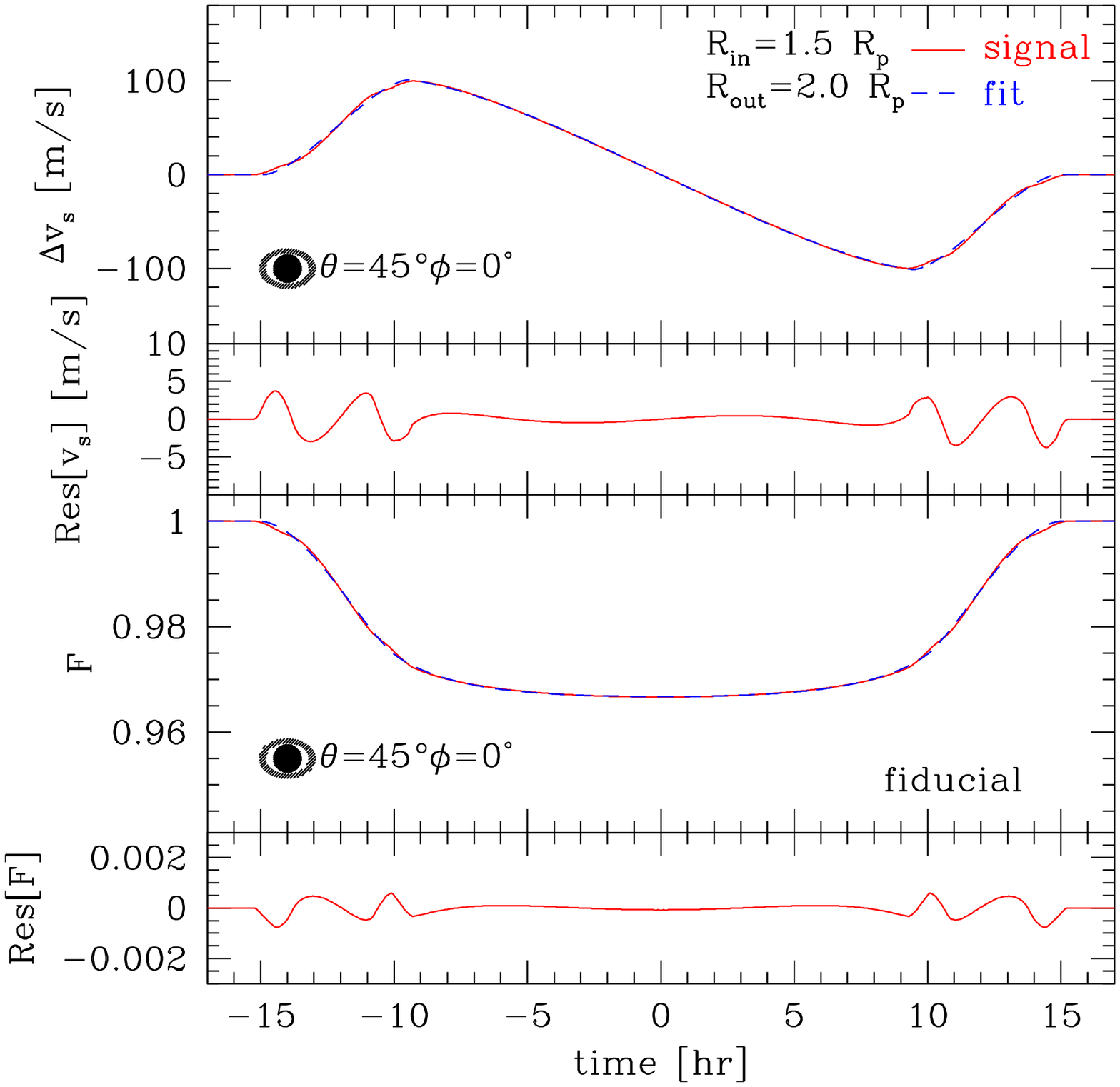}{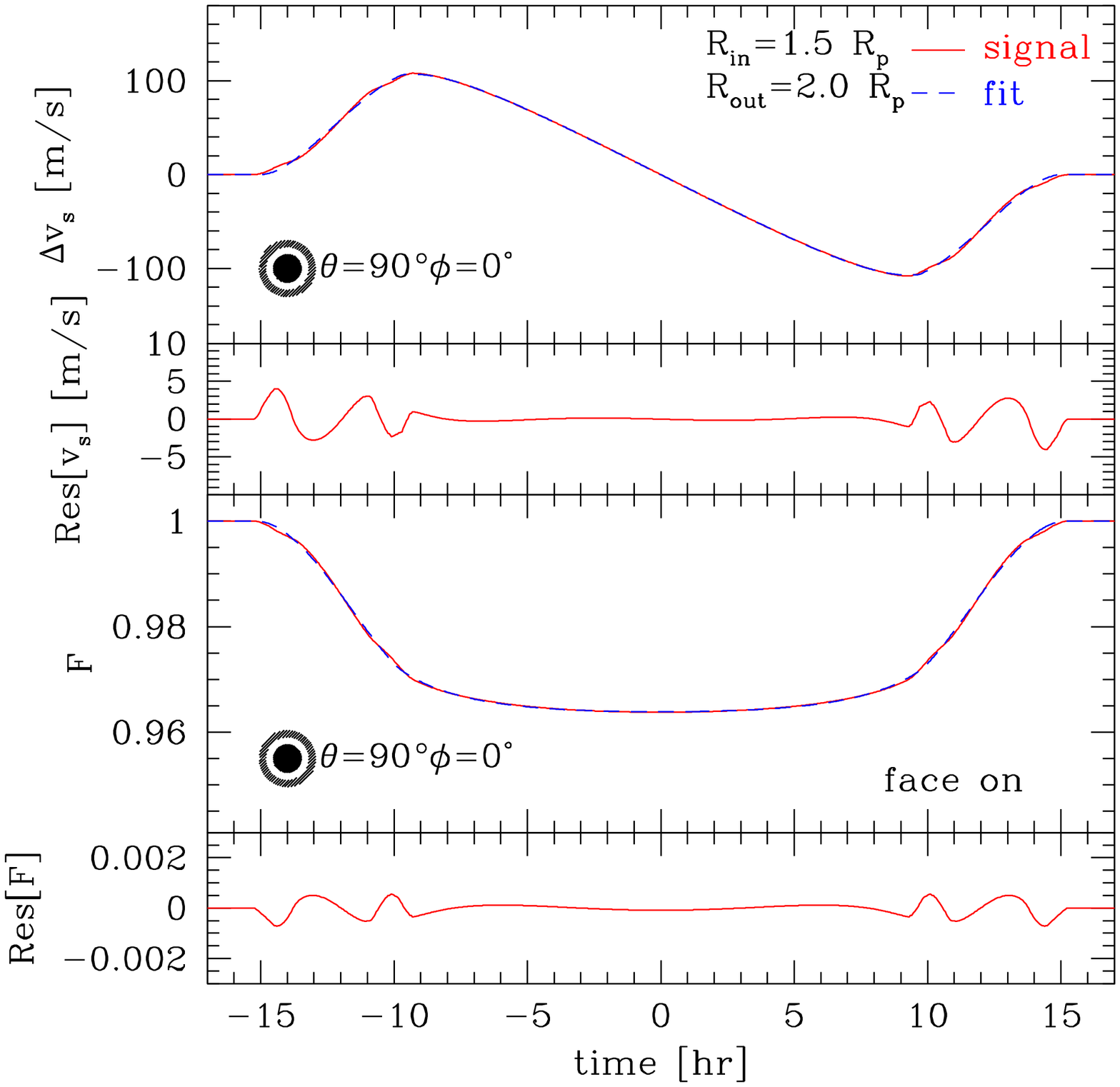}\\
 \plottwo{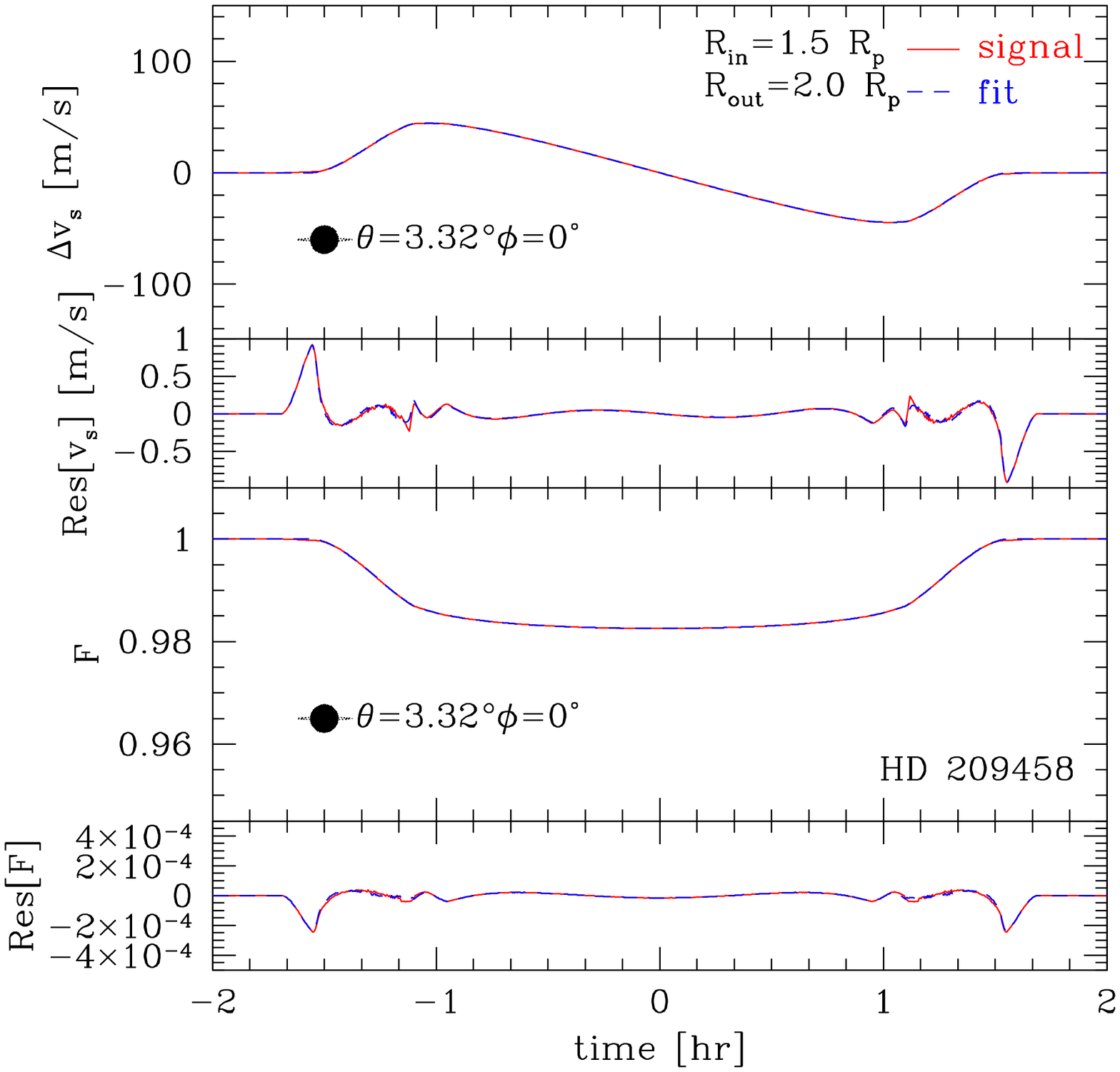}{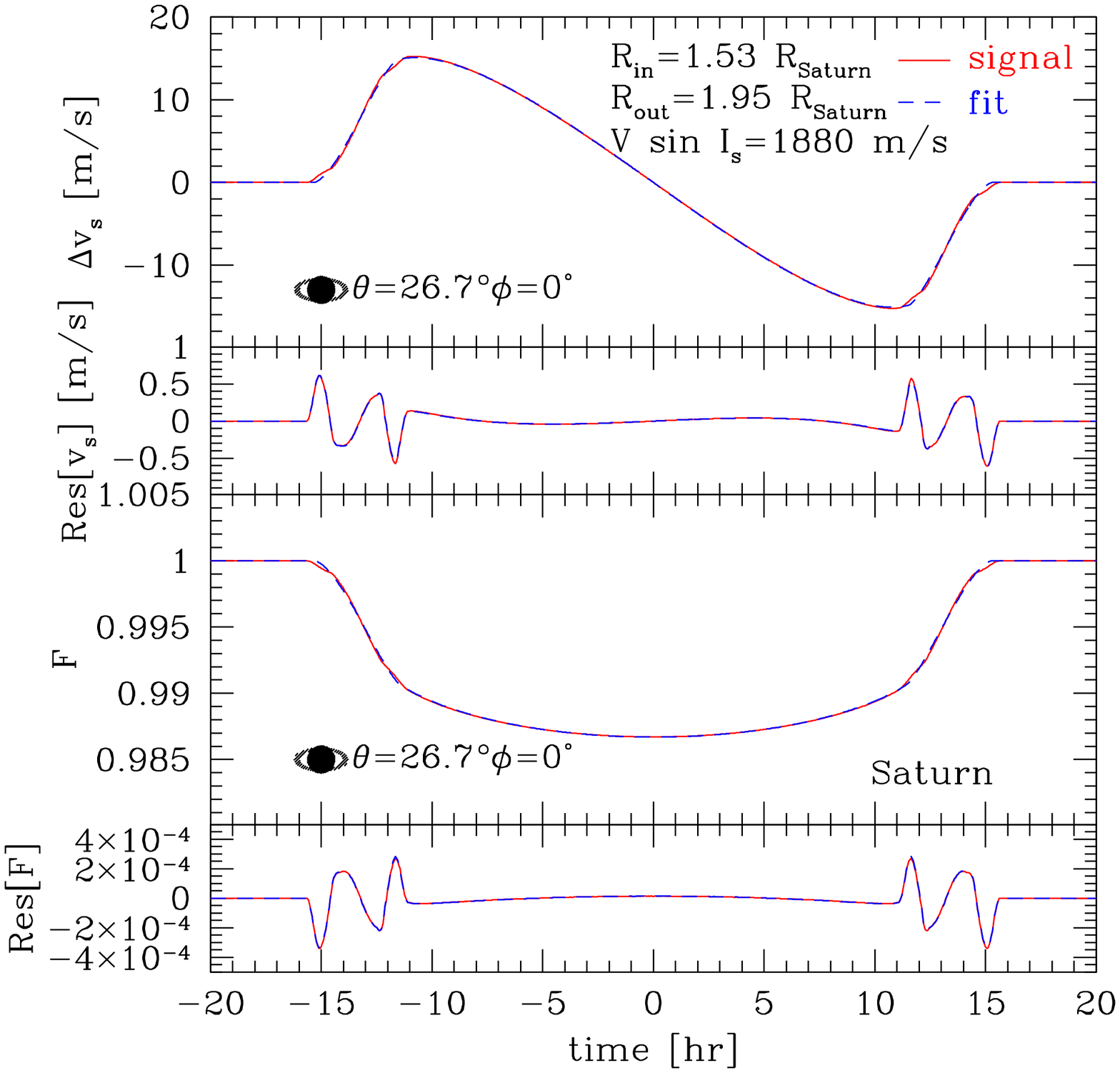}
\figcaption{Spectroscopic and photometric signatures of a ring.  Each
panel displays the velocity anomaly due to the Rossiter effect $\Delta
v_s$, the spectroscopic residual $\mbox{Res}[v_s]$, photometric
light curve $F$, and the photometric residual $\mbox{Res}[F]$ from top to
bottom. Solid and dashed curves in $\Delta v_s$ and $F$ present our
predictions and the best-fit without ring model. 
Note that in lower panels, 
the residuals obtained from the numerical fit 
are also plotted as dashed lines, 
in addition to the one obtained from the analytical fit ({\it solid})
; $\theta=45^\circ, \phi=0^\circ$ ({\it upper-left}), $\theta=90^\circ,
\phi=0^\circ$ ({\it upper-right}), tidally-locked close-in planet 
adopting the orbital parameters of HD 209458 system, i.e., 
$a=0.0468$AU, $\theta=3^\circ.32, \phi=0^\circ$ ({\it lower-left}),  
and Saturn system (B-ring) with
$\theta=26^\circ.7, \phi=0^\circ$, $R_{\rm in}=1.53 R_{\rm Saturn}$, 
$R_{\rm out}=1.95 R_{\rm Saturn}$, $V\sin I_s = 1,880\, {\rm m}\,
{\rm s}^{-1}$ ({\it lower-right}).  \label{fig:fiducial}}
\end{figure}

\citet{Lovis06} has already achieved $\sim 1$m\,s$^{-1}$ sensitivity in
the radial velocity measurement.  Also the HST photometric accuracy is
shown to be better than $0.1$\% \citep{Brown01}, which is expected to be
the case for upcoming space missions as well.  Figure \ref{fig:fiducial}
illustrates examples for $\mbox{Res}[v_s]$ and $\mbox{Res}[F]$; the
upper and lower-left panels show our fiducial ring model (Table
\ref{tbl:parameters}) with three different orientations,
($\theta=45^\circ, \phi=0^\circ$), ($\theta=90^\circ, \phi=0^\circ$) and
($\theta=90^\circ-i_{\rm orbit}=3^\circ.32, \phi=0^\circ$).  The last
case corresponds to a close-in planet, HD 209458b, with ring.  The
semi-major axis and inclination angle of the orbital plane with respect
to our line-of-sight are set to $a=0.0468$AU and $i_{\rm
orbit}=86^\circ.68$, respectively. Since close-in planet is likely to be
a tidally-locked system, we assume that the ring plane lies in the
planetary orbital plane. Thus, $\theta$ is equal to $90^{\circ}-i_{\rm
orbit}$. The lower-right panel is intended to illustrate the expected
signals for the Saturn system (B-ring) just for comparison;
$\theta=26^\circ.7, \phi=0^\circ$, $R_{\rm in}=1.53 R_{\rm Saturn}$,
$R_{\rm out}=1.95 R_{\rm Saturn}$, $V\sin I_s =1,880 {\rm m}\,{\rm
s}^{-1}$.  Note that in the lower panels, the residuals obtained from
the numerical fit are also plotted as dashed lines, in addition to the
results from the analytical fit ({\it solid}).  The residuals,
$\mbox{Res}[v_s]$ and $\mbox{Res}[F]$, clearly indicate the
characteristic pattern expected for a ring. More importantly, the above
signatures are indeed at a few $\sigma$ level of the currently best
observational sensitivities, both photometrically and spectroscopically.
Even though it is not clear if those values that we adopted for the ring
parameters are realistic, the result is certainly encouraging.

\begin{figure}[htp]
 \epsscale{1.1}
 \plottwo{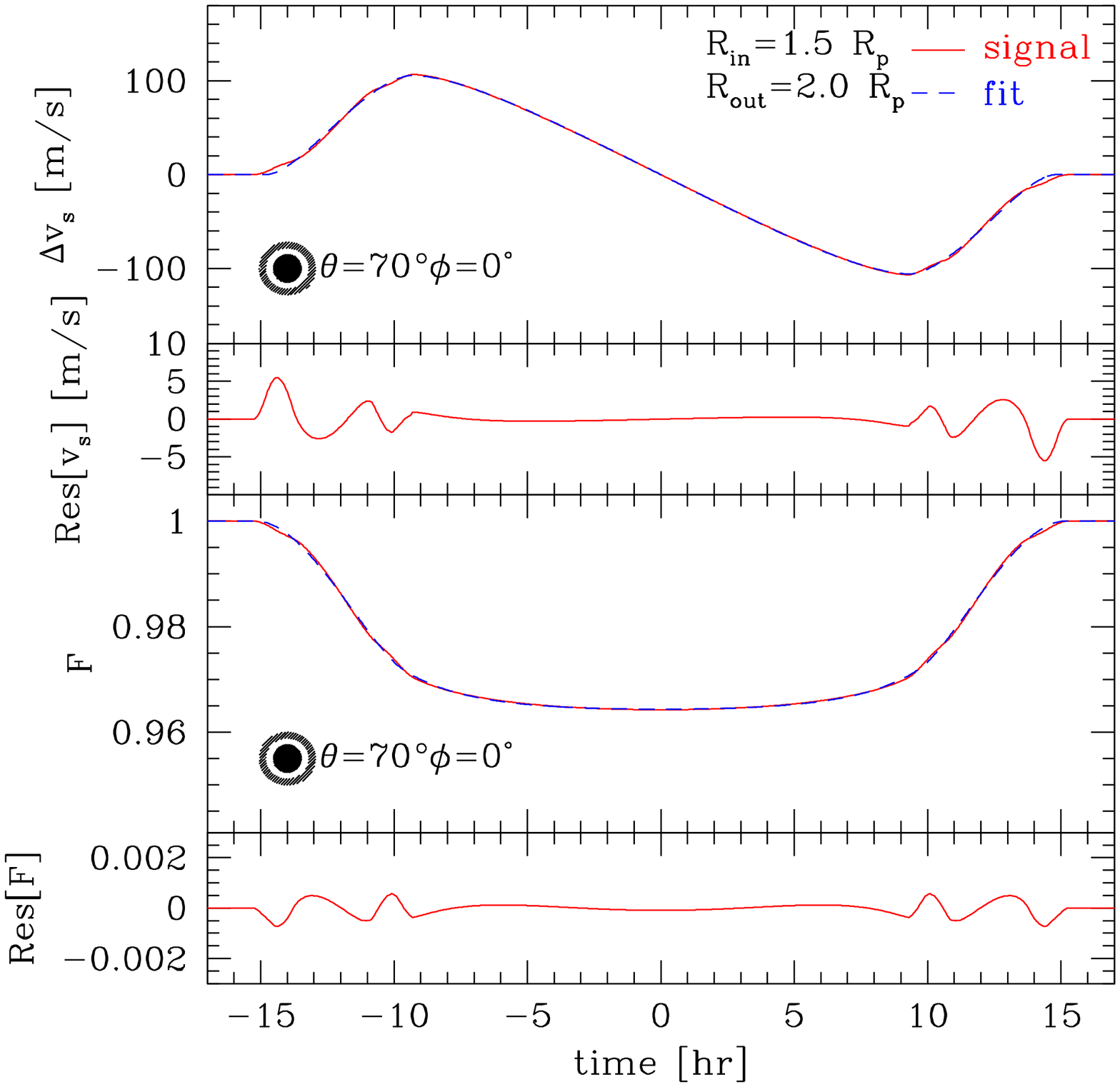}{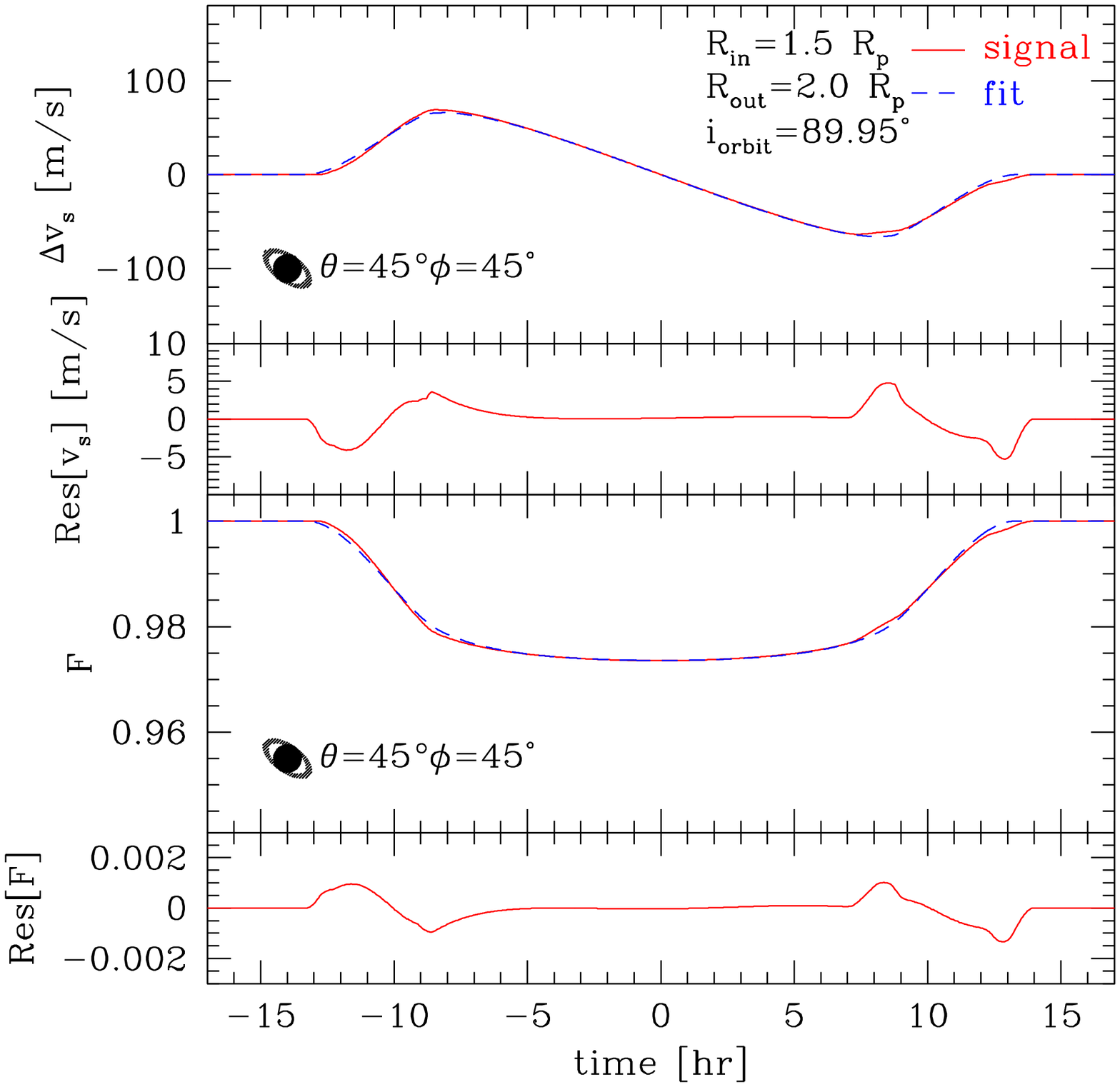}\\
 \plottwo{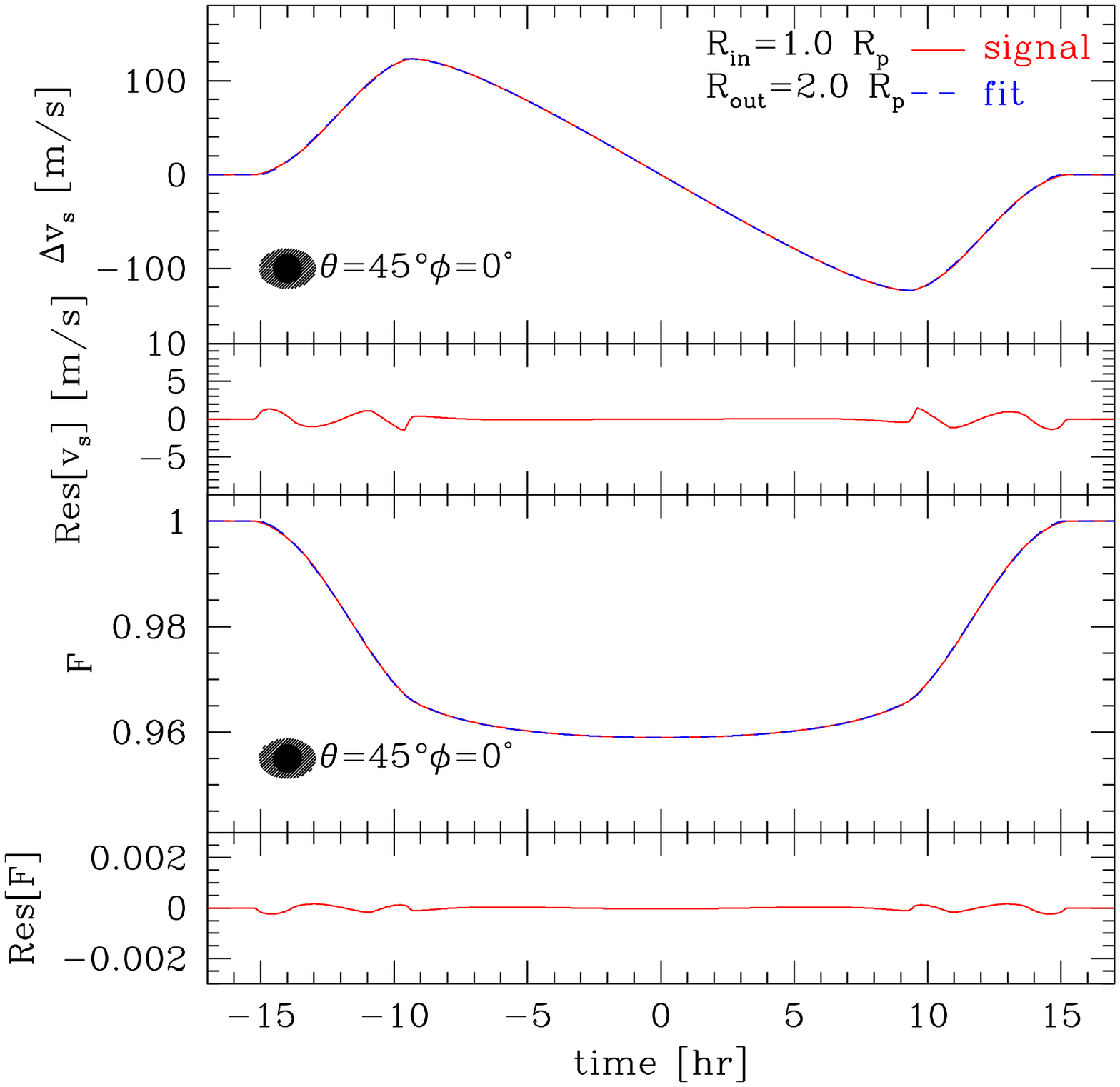}{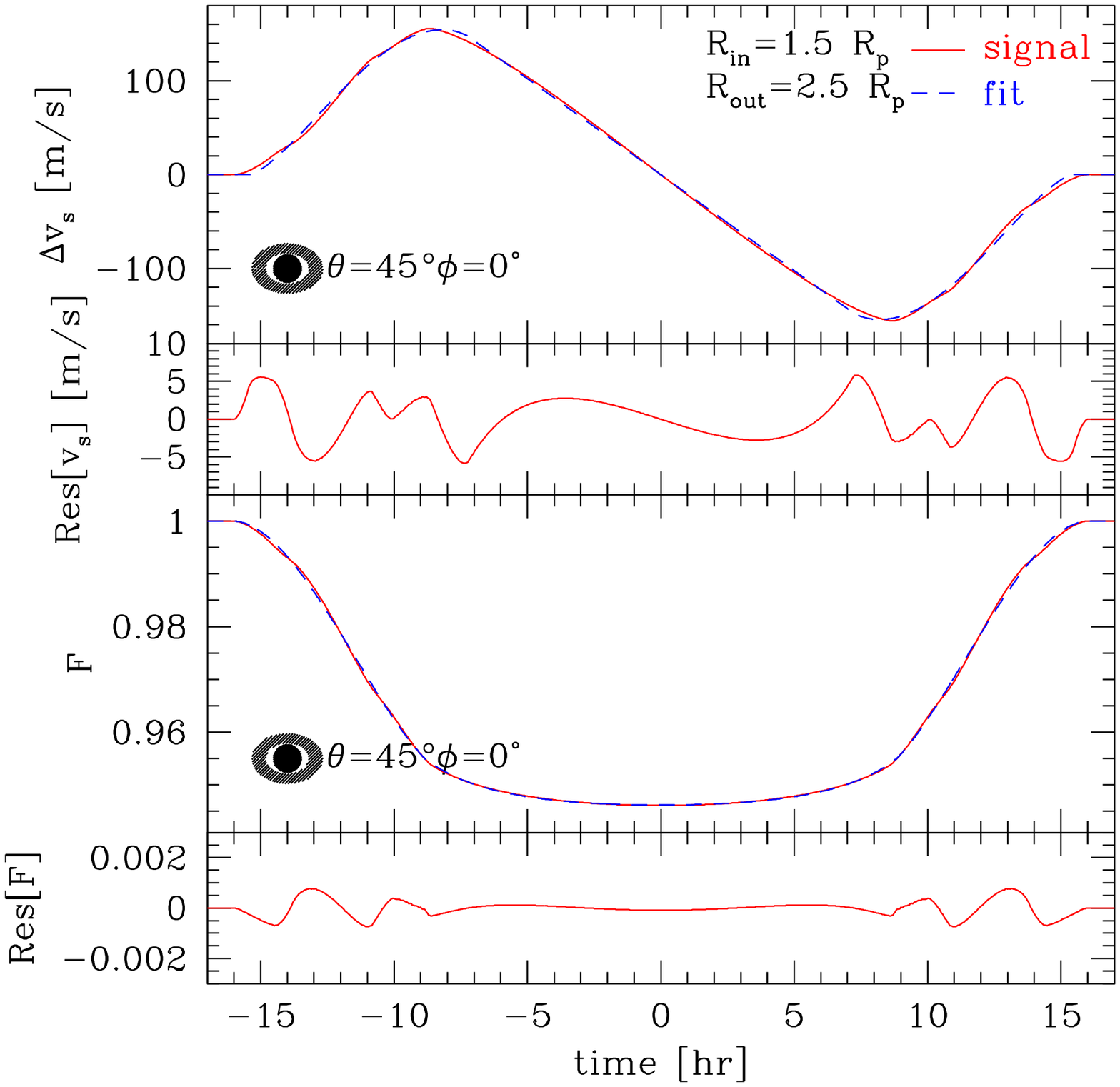}
\figcaption{ Same as the upper-left panel of Figure \ref{fig:fiducial}, 
but with a different parameter set; $\theta=70^\circ$ ({\it upper-left}),  
$\phi=45^\circ$ and $i_{\rm orbit}=89^\circ.95$ ({\it upper-right}), and 
$R_{\rm in}=1R_p$ ({\it lower-left}) 
and $R_{\rm out}=2.5R_p$ ({\it lower-right}).  
\label{fig:fit ring}}
\end{figure}

Figure \ref{fig:fit ring} displays the results when one parameter of the
ring system is changed from the fiducial values, $\theta=45^\circ,
\phi=0^\circ$, $R_{\rm in}=1.5 R_{\rm p}$, and $R_{\rm out}=2.0 R_{\rm
p}$ (the upper-left panel of Fig.\ref{fig:fiducial}); $\theta=70^\circ$
({\it upper-left}), $\phi=45^\circ$ and $i_{\rm orbit}=89^\circ.95$({\it upper-right}), and $R_{\rm
in}=1.0 R_{\rm p}$ ({\it lower-left}) and $R_{\rm out}=2.5 R_{\rm p}$
({\it lower-right}).  The behavior is fairly easy to understand in an
intuitive manner. As long as the clear gap between the planetary surface
and the inner edge of the ring exists, a ring produces a larger residual
as $\theta$ approaches $90^{\circ}$ (face-on); an edge-on system is hard
to identify. A non-zero value of $\phi$ generates an asymmetry in the
residuals depending on the value of the inclination angle of the orbit
$i_{\rm orbit}$; $\mbox{Res}[F]$ is not symmetric with respect to $t
\leftrightarrow -t$, and $\mbox{Res}[v_s]$ loses a symmetry with respect
to the origin.  The latter effect partially resembles the behavior of
the spin-orbit misalignment and results in a systematic offset of the
value of $\lambda$.  While we restrict our analysis here for $\lambda=
0^{\circ}$ case, this needs to be kept in mind in future precise
analyses of ring systems in general.  If the width of the ring, $\Delta
R=R_{\rm out}-R_{\rm in}$, is larger, the signals themselves always
increase, but the residuals may decrease in some cases (lower-left panel
of Fig.\ref{fig:fit ring}). A trivial example is a face-on ring system
with $R_{\rm in}=R_p$ which cannot be distinguished from a single planet
if the ring is opaque.  The gap between the planet and the ring leaves
an important observational signature of the ring.

Incidentally \citet{Colwell07} showed that the line-of-sight optical
depth of Saturn's ring is roughly independent of the viewing angle.  If
this is the case, we should set $\beta=1$ in equation
(\ref{eq:intensity_I}). Thus we repeated the same calculations for a
variety of different values of $\theta$ by setting $\beta=1$ instead.
We found that the amplitudes of the {\it observable} ring signatures,
i.e., residuals defined by equations (\ref{eq:Res_v_s}) and
(\ref{eq:Res_F}) are not so sensitive to the assumption for the viewing
angle dependence of $\tau$. Since our current procedure looks for the
difference with respect to the best-fit ringless model, a part of the
true signals of the ring is necessarily lost by being misidentified as a
contribution of the planet. This is why the residual signatures of rings
shown in Figures \ref{fig:fiducial} and \ref{fig:fit ring} are fairly
robust against the assumption for the viewing angle dependence of
$\tau$.

\subsection{Detectability of rings \label{subsec:detectability}}

So far we have ignored observational errors of instrumental noises.  In
this subsection, we discuss the detectability of the ring system and
clarify under what conditions the presence of rings is statistically
confirmed and is really established as an undisputed fact.  For this
purpose, we create mock data and measure the residuals of $\Delta v_s$
and $F$. Using 1,000 mock realizations, the statistical significance of
the residuals is estimated.

We define 
\begin{equation}
\chi^2 = \sum_{i=1}^{N_{\mathcal O}} \left[\frac{\mbox{Res}[\mathcal{O}_i]}
{\sigma_{\mathcal O}}\right]^2 
\label{eq:def_of_chi2}
\end{equation}
with the subscript $i$ stands for the sample point. Here, the quantity
${\mathcal O}$ denotes either $\Delta v_s$ or $F$; $\sigma_{\Delta v_s}$
and $\sigma_F$ are the statistical errors of radial velocity anomaly and
relative flux ratio, respectively. We specifically assign the errors of
$\sigma_{\Delta v_s}=1$\,m~s$^{-1}$ and $\sigma_F=10^{-4}$ as currently
achieved precision levels, and also of $\sigma_{\Delta
v_s}=0.1$\,m~s$^{-1}$ and $\sigma_F=10^{-5}$ as future precision. The
mock data are then created adopting the central values of the fiducial
model in Table \ref{tbl:parameters} with Gaussian random errors of
$\sigma_{\Delta v_s}$ and $\sigma_F$.

\begin{figure}[htbp]
 \epsscale{1.1}
 \plottwo{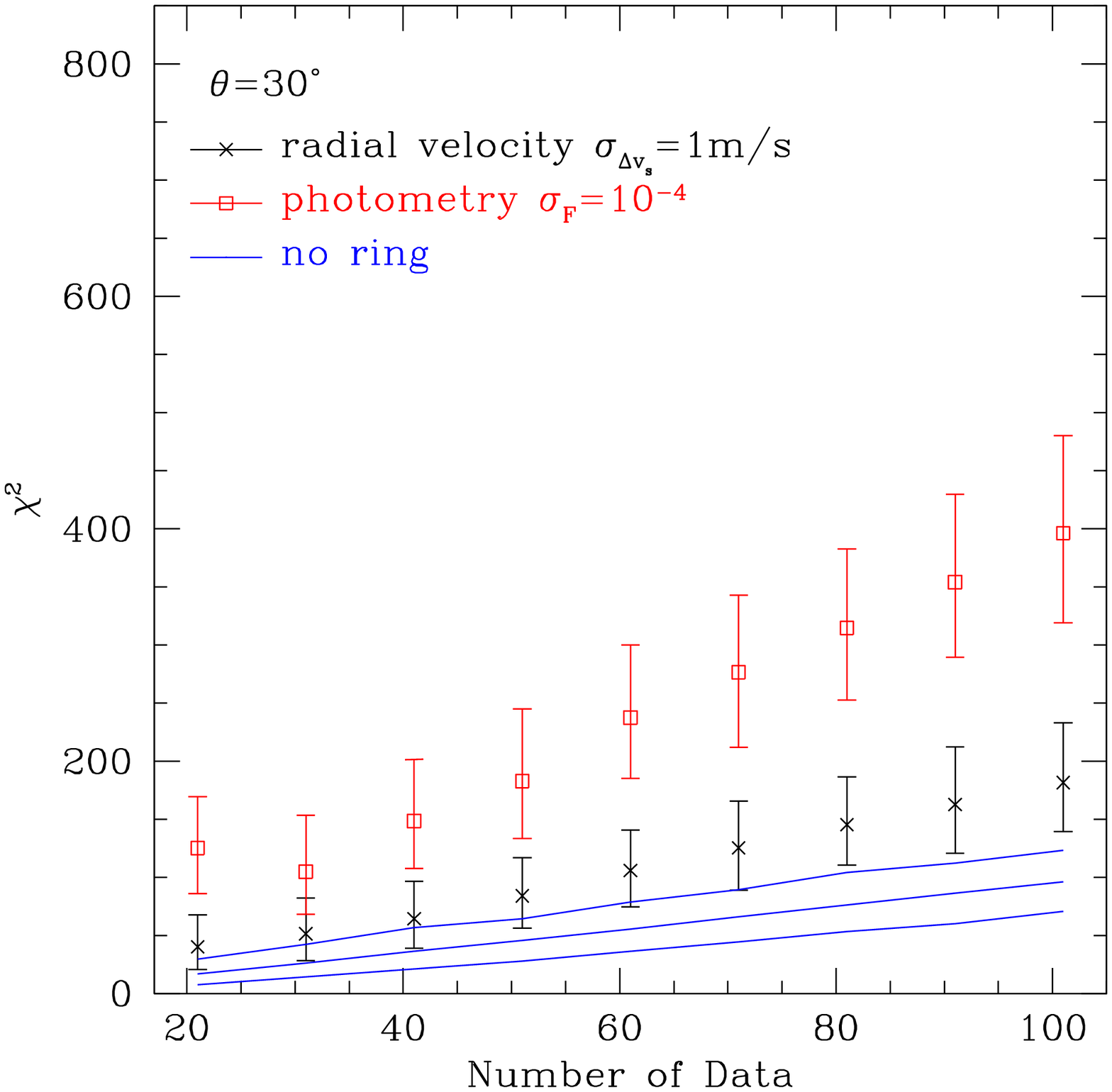}{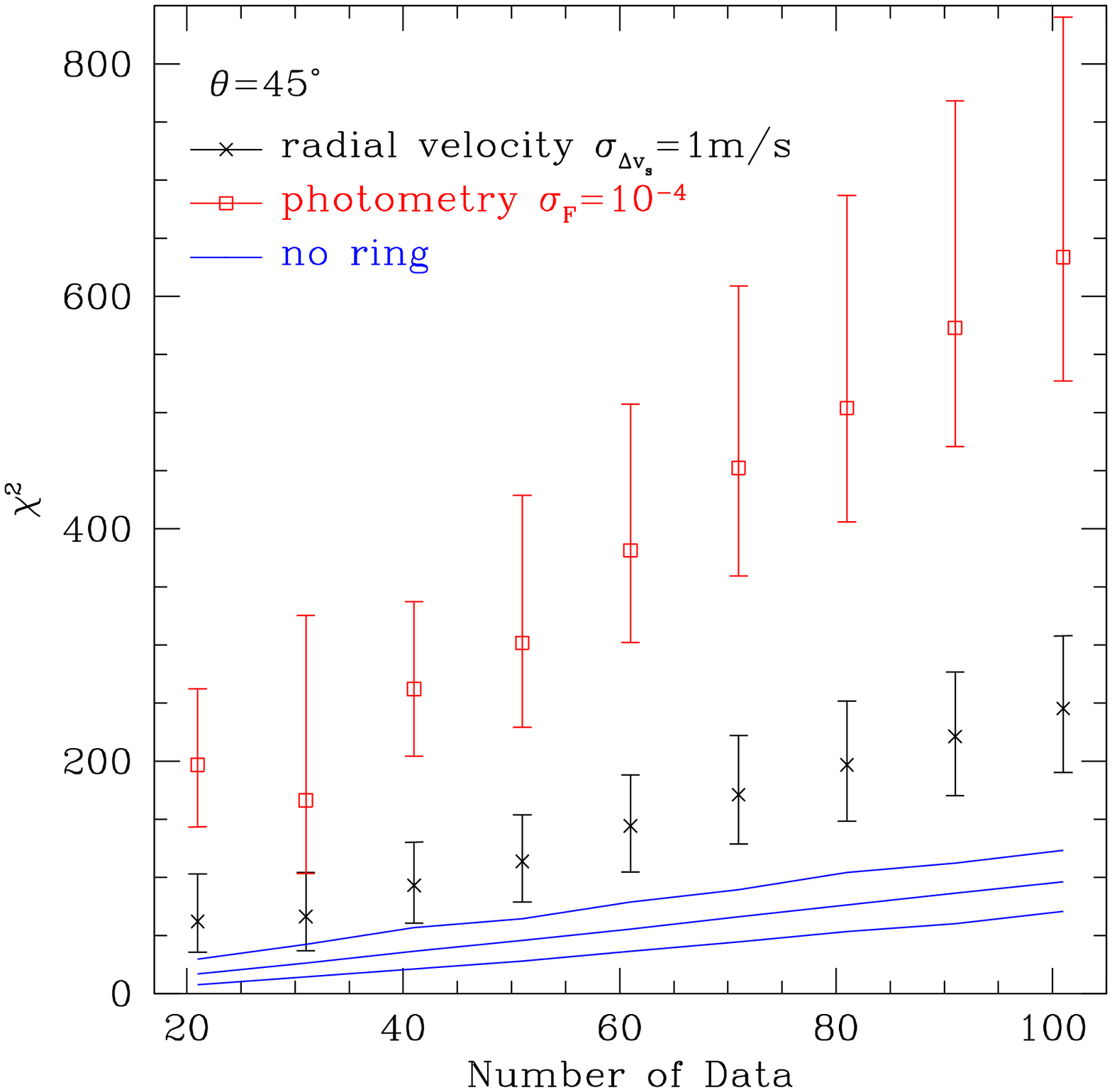}
 \epsscale{0.5}
 \plotone{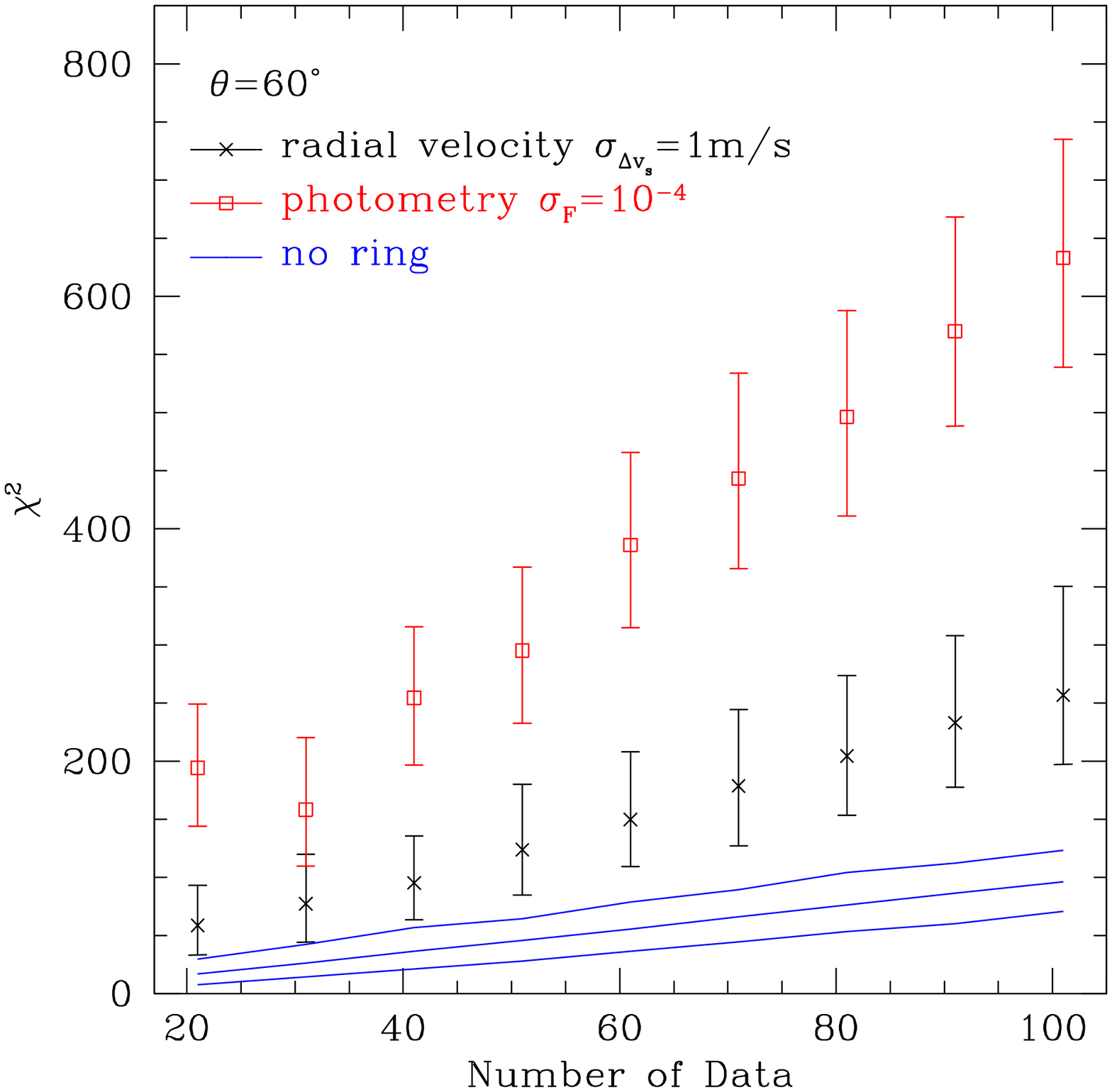}
 \figcaption{$\chi^2$ with respect to the the best-fit parameters for a
ringless planet model against the number of sampled data points
(eq.[\ref{eq:def_of_chi2}]); the mean value of $\chi^2$ are plotted in
symbols with the error bars indicating the 95\% confidence interval
estimated from 1,000 mock realizations.  Open squares and crosses denote
$\chi^2$ for photometric data and for spectroscopic data, respectively
The mock data were created adopting the fiducial values listed in Table
\ref{tbl:parameters}, except for the obliquities of the ring axis;
$\theta=30^\circ$ (upper-left), $45^\circ$ (upper-right), and
$\theta=60^\circ$ (lower).  
The spectroscopic and photometric
accuracies of ($1$ m s$^{-1}$, $10^{-4}$) are adopted.
\label{fig:scatter_likelihood}}
\end{figure}

In Figure \ref{fig:scatter_likelihood}, we estimate $\chi^2$ from 1,000
mock realizations and quantify the scatter of the $\chi^2$ functions by
varying the number of data samples and the ring parameters of mock data.
Basically, we choose the central transiting time as the origin of the time
and sample the data at regular interval. For the cases with future 
precision (bottom panels of Fig.9), however, we shift the origin of the 
binning randomly within one bin size with respect to the central transiting
time. We then explore the effects of bin size by varying the number of
data points and computing $\chi^2$.  If we do not randomly shift the
origin of the binning, $\chi^2$ shows an artificial oscillation due to
the sinusoidal nature of the expected signal, and this becomes prominent 
for the cases with higher precision. 
In order to avoid the artificial coherence of the sampled phase, 
we therefore shift the central position
by a size that is randomly chosen between 0 to the size of bin.

Each panel in Figure \ref{fig:scatter_likelihood} shows the 95\%
confidence interval for the distribution of $\chi^2$ values around the
mean values ({\it open squares} for photometric measurements, {\it
crosses} for spectroscopic measurement). As a reference, we also
estimated the scatter of $\chi^2$ values from the mock data without
rings, and the resulting 95\% confidence interval are plotted as solid
lines, with middle lines being mean values. Hence, significant deviation
from the three reference lines disfavors a ringless planet, suggesting
the presence of ring as a plausible solution. In each panel, the mock
data were created adopting the fiducial values listed in Table
\ref{tbl:parameters}, except for the obliquity of ring parameters,
$\theta$.  We specifically examine the three cases: $\theta=45^{\circ}$
(upper-left), $30^{\circ}$ (upper-right), and $60^{\circ}$ (lower).

Figure \ref{fig:scatter_likelihood} indicates that rings with a large
obliquity are easily detected. The increased number of both the
photometric and spectroscopic samples also improves the detectability,
although the photometric data tend to have a better sensitivity compared
to the spectroscopic data. This is partly because characterization of the
spectroscopic signal requires two additional parameters, i.e., $V\sin
I_s$ and $\lambda$, in addition to
the model parameters of photometric signals.
Thus, $\chi^2$ values of the spectroscopic measurement are expected to be
generally small. Nevertheless, spectroscopic detection/confirmation of
rings is valuable and combined data analysis will significantly improve
the reliability of the detection. This is especially relevant for
planets orbiting around a rapidly rotating star, in which case the
spectroscopic signatures of ring during ingress and egress phase become
more prominent.

\begin{figure}[hp]
 \epsscale{1.1}
 \plottwo{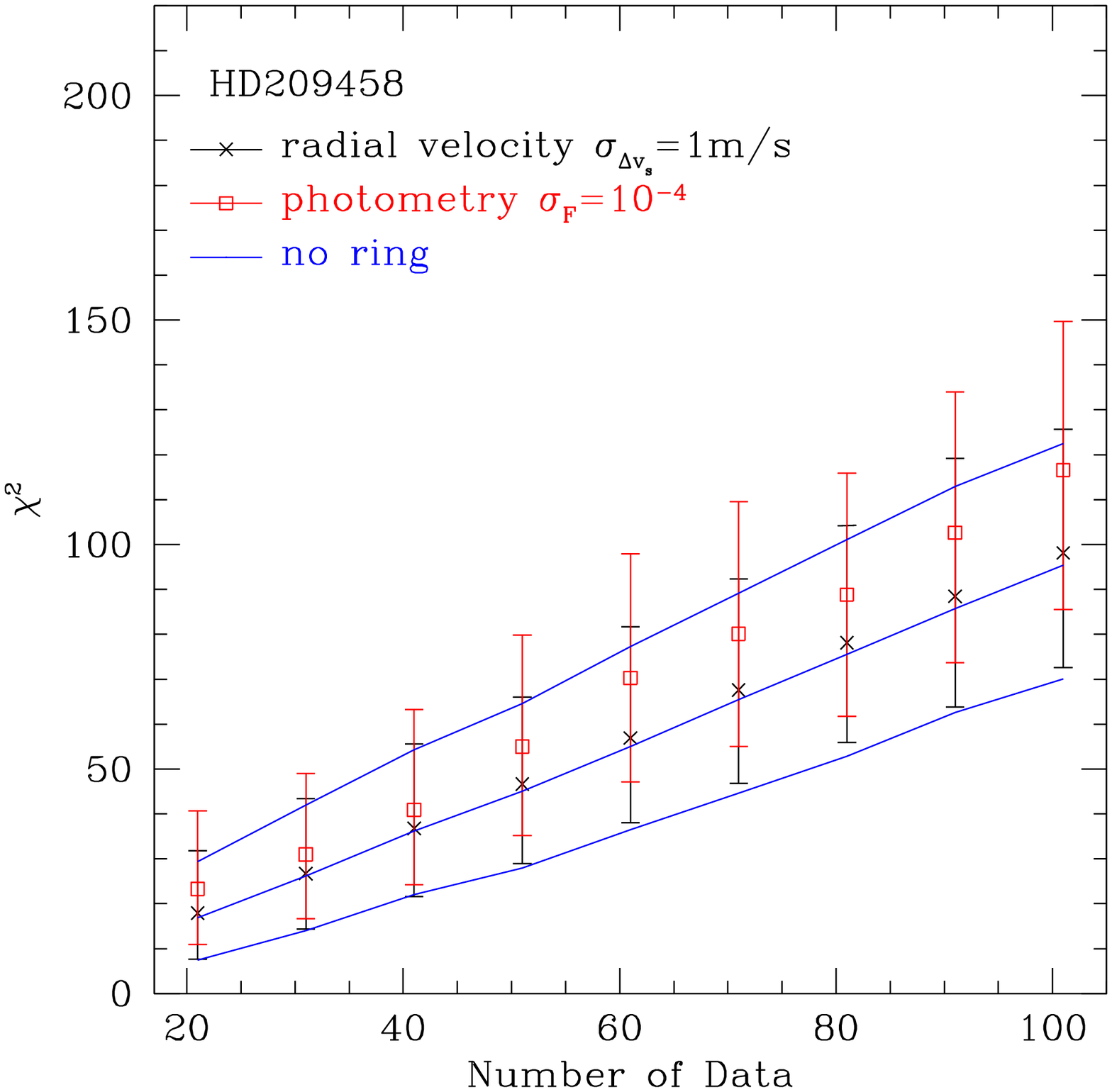}{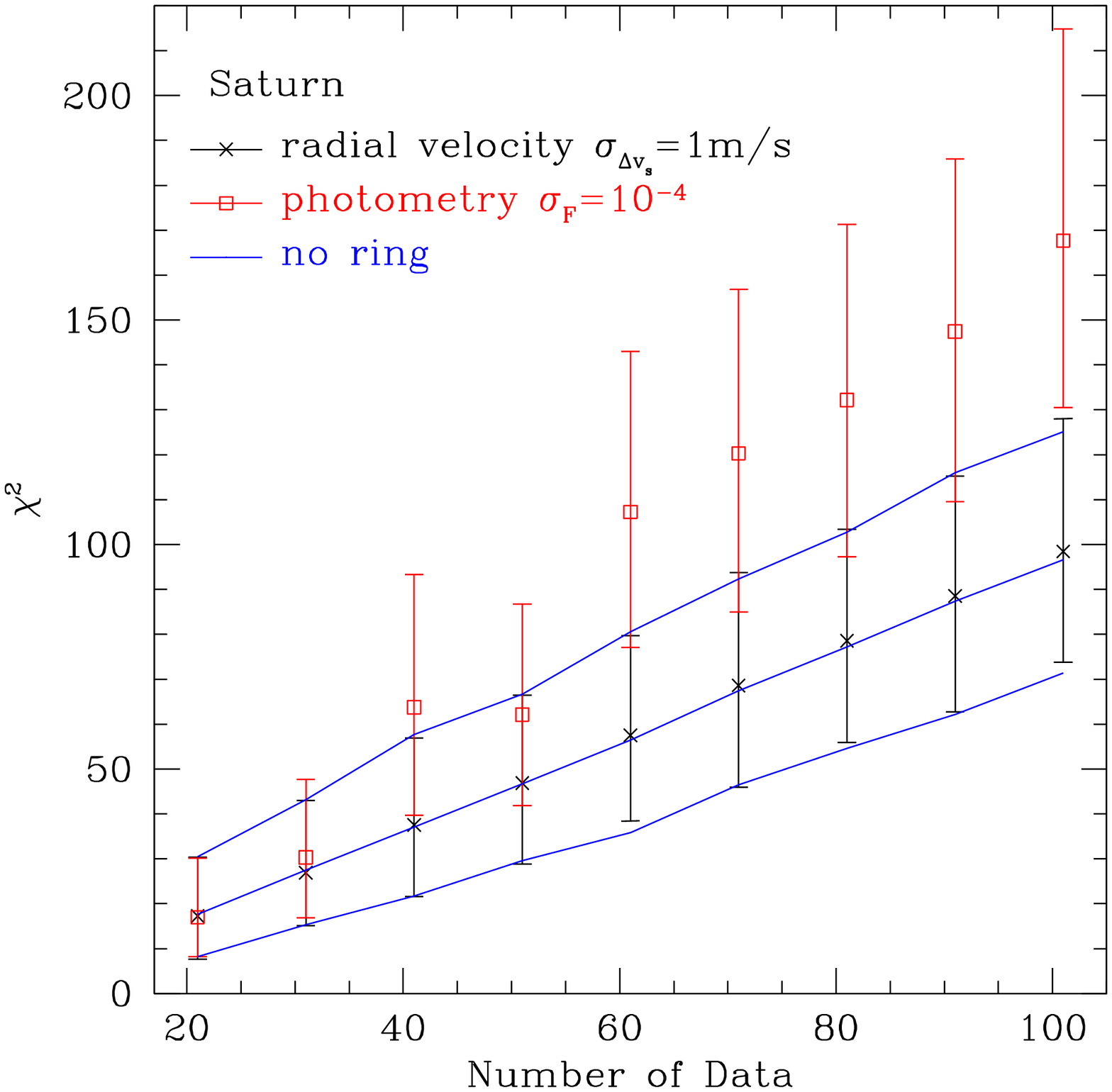}
 \plottwo{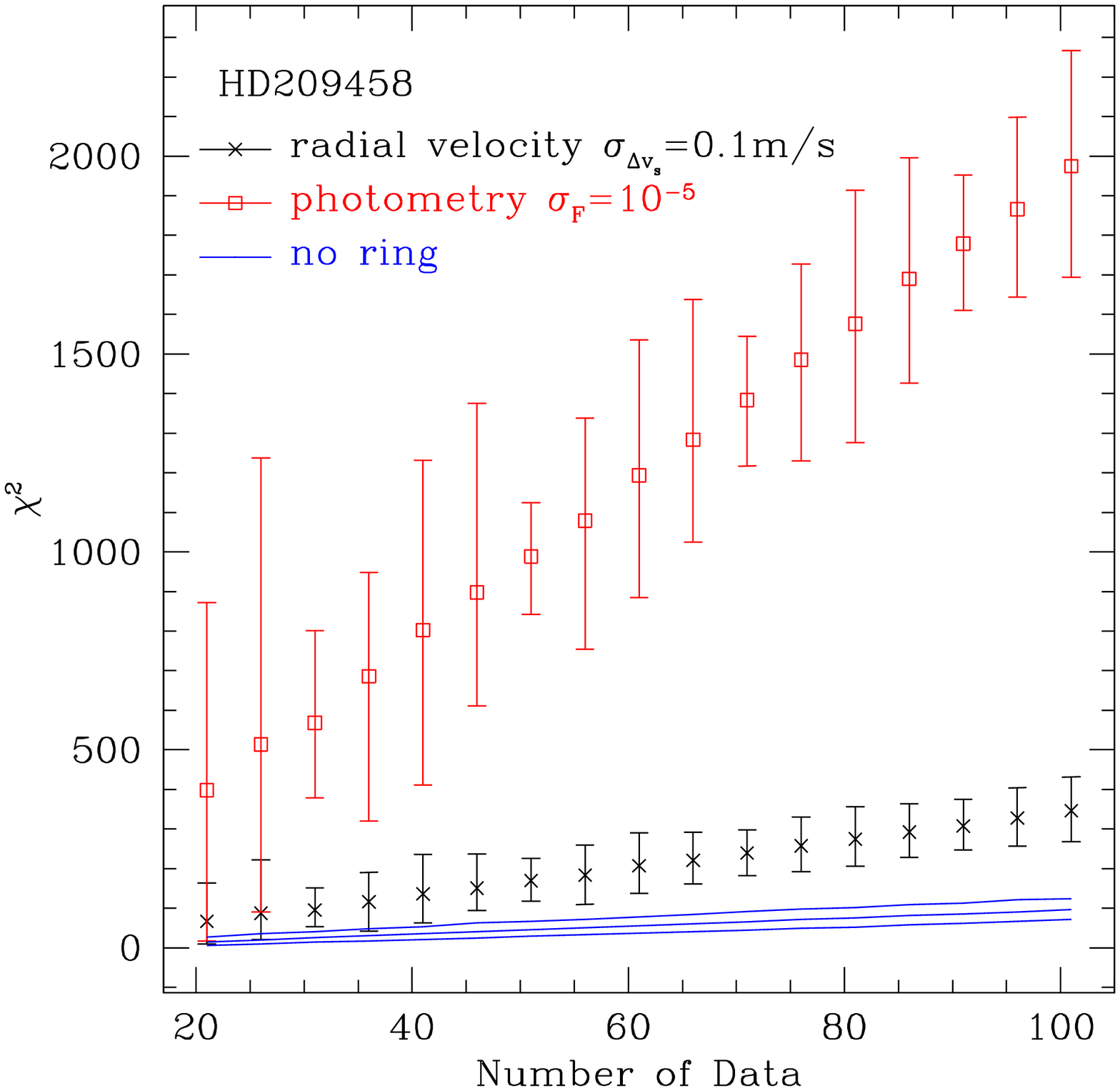}{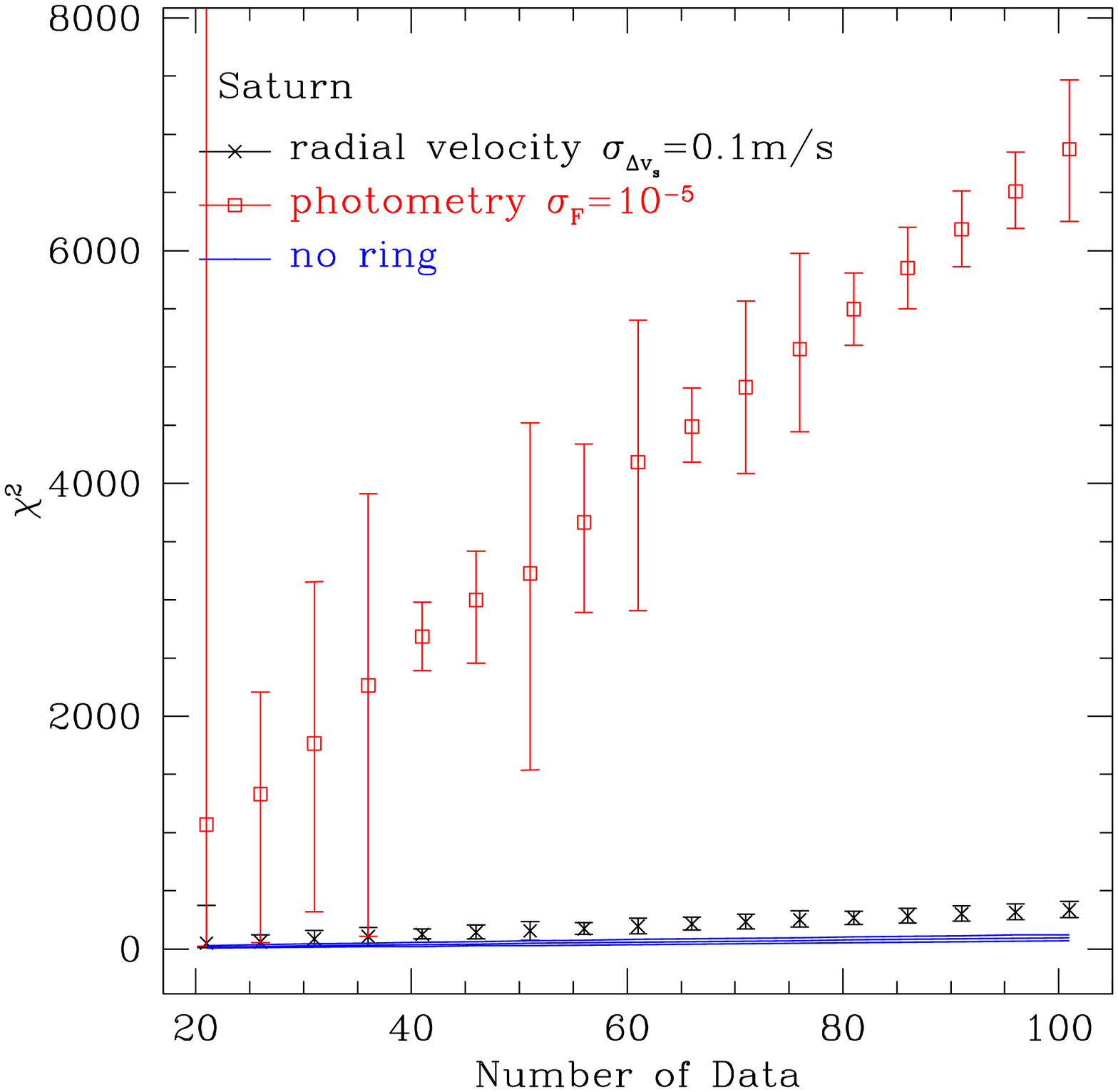}
\figcaption{Same as Figure \ref{fig:scatter_likelihood}, but for 
HD 209458 and Saturn systems. Left panels show the results for 
a tidally-locked close-in planet adopting the orbital parameters of 
HD 209458 system, while right panels are for Saturn system. 
The model parameters of these examples are described in caption of 
Figure \ref{fig:fiducial}.
The upper and lower panels adopt the spectroscopic and photometric
 accuracies of ($1$ m s$^{-1}$, $10^{-4}$) and ($0.1$ m s$^{-1}$,
 $10^{-5}$), respectively.
\label{fig:scatter_likelihood2} }
\end{figure}

In Figure \ref{fig:scatter_likelihood2}, we also examine the cases for
the close-in planet (HD 209458b) assuming a tidally-locked system ({\it
left panels}) and Saturn system ({\it right panels}).  The close-in
planet is likely to be a tidally-locked system and the obliquity of the
ring is thus expected to be rather small in such a case.  Thus a
possible signature of a ring is inevitably weak and it is difficult to
detect with the current sensitivity (Figure \ref{fig:fiducial}). On the
other hand, Saturn has a relatively large obliquity
($\theta=26.7^\circ$) and the photometric signatures of the ring is
marginally detectable with the HST photometric accuracy of $10^{-4}$.
The spectroscopic signatures are a bit below the detection level of 1 m
s$^{-1}$ partly because the rotation velocity of the Sun is small
($V\sin I_s=1,880$m\,s$^{-1}$).  

Nevertheless, the improved precision of future observations will
significantly increase the detection threshold. Indeed, the accuracy of
spectroscopic measurement is supposed to achieve $\sim$ a few
cm\,s$^{-1}$ level \citep[][]{Li_etal08,Murphy_etal07}, and Kepler may
reach $\sigma_{\rm F} \approx 10^{-5}$. The lower panels of Figure
\ref{fig:scatter_likelihood2} assume those future precisions.  They
indicate that if possible systematic effects due to the stellar activity
are well under control, the spectroscopic signatures of rings from the
ground-based observation are even detectable.

\section{Conclusions and discussion}

We have explored photometric and spectroscopic detectability of rings
around extrasolar transiting planets. We have described a general
formulation to predict both the photometric transiting signature in the
stellar light curve and the spectroscopic radial velocity anomaly due to
the Rossiter effect for a planetary system with a ring. For a face-on
ring system, we have provided analytic formulae for those effects, which
correspond to an improved version of our previous results (Paper I).  We
have shown also several numerical results for more general orientations
of planetary rings. We found that possible planetary rings around
HD 209458b and Saturn may be marginally detectable even with the current
observational accuracy as long as sufficiently sampled data points are
available. This is very encouraging, and we believe that the detection
of rings around extrasolar transiting planets is one of the realistic
scientific goals of up-coming space missions.

Let us make a couple of additional comments here. First, our present
results may be directly applicable to oblate planets (by setting
$\tau=\infty$, $R_{\rm p}=R_{\rm in}=0$, and regarding $R_{\rm out}$ as
the surface of the oblate planet, for instance). It would be unlikely
that the slow spin velocity of tidally-locked close-in planets
significantly distorts their surface, but it would be interesting to put
any constraints on the oblateness of the planets
\citep[e.g.,][]{Seager02}.  Second, while one would expect that the
planetary spin axis is aligned with the orbital axis, \citet{WH05}
argued an interesting possibility that this is not the case. Thus the
detection of the ring can test their proposal in a straightforward
manner if the planetary spin axis is perpendicular to the ring.

We have to admit, however, that the existence of a Saturnian planet with
a ring in extra-solar systems is very speculative, in particular for
close-in planets where any interpolation from our Solar system may not
be justified. Thus we share the same feeling that Nagaoka expressed
more than a hundred years ago even in the historical paper
\citep{nagaoka04}: {\it ``There are various problems which will possibly
be capable of being attacked on the hypothesis of a Saturnian system
''}. Nevertheless we cannot resist the temptation to conclude our paper
by quoting the last sentence in his paper: {\it ``The rough calculation
and rather unpolished exposition of various phenomena above sketched may
serve as a hint to a more complete solution of atomic structure.''}

Finally it would be interesting to note the analogy between the
extrasolar planetary system and the atomic model; radial velocity
modulation of stars due to planets, the Rossiter effect of transiting
planets, and the signature of rings correspond to the atomic orbital
angular momentum, the spin of proton, and the spin of electrons.  We
should not forget that the wild idea of Nagaoka led to quantum physics,
which turned out to provide the physical basis of the astronomical
spectroscopic observations via the orbital and spin angular momenta of
various atomic systems, and therefore he indeed opened a way to the
extrasolar planet search.

\vspace*{0.5cm}

We thank Joshua Winn, Edwin Turner and Norio Narita for useful
discussions, and Erik Reese for a careful reading of the manuscript. We
also thank an anonymous referee for several constructive comments which
improved the presentation of the earlier manuscript. This work is
partially supported by Grants-in-Aid for Scientific Research from the
Japan Society for Promotion of Science (No.14102004, 16340053 and
18740132).

\appendix
\section{Approximation of Integrals \label{appendix:formula}}

Here, we present the approximate expressions for the integrals $W_i$
$(i=1$--$4)$ that appear in the formulae (\ref{eq:face-on_1}) and
(\ref{eq:face-on_2}). These quantities are all expressed in terms of the
planet position $\rho=(X_p^2+Z_p^2)^{1/2}$ together with the small
parameter $\gamps \equiv R_p/R_s$.  Before writing down the approximate
formulae, we give precise definitions of these integrals (see Sec.5 of
Paper I):
\begin{eqnarray}
W_1&=& \frac{1}{\pi R_p^2}\, \iint_S dxdz \sqrt{1-(x^2+z^2)/R_s^2},
\label{appen:W_1_def}
\\
W_2&=& \frac{1}{X_p\, \pi R_p^2}\,
\iint_S dx dz\,\, x\sqrt{1-(x^2+z^2)/R_s^2},
\label{appen:W_2_def}
\\
W_3 &=& \iint_{S'} d\tilde{x}d\tilde{z} \sqrt{1-\tilde{x}^2-\tilde{z}^2},
\label{appen:W_3} 
\\
W_4 &=& \iint_{S'} d\tilde{x}d\tilde{z} \,\,\tilde{x}
\sqrt{1-\tilde{x}^2-\tilde{z}^2}, 
\label{appen:W_4} 
\end{eqnarray}
where the variables $(x,z)$ means the coordinate system defined in Figure 
\ref{fig:orbit}. The variables $(\tilde{x},\tilde{z})$ are related to 
the coordinate $(x,z)$ through 
\begin{eqnarray}
 \left(
\begin{array}{c} 
\tilde{x}\\ 
\tilde{z} 
\end{array}
\right) =  \frac{1}{R_s\sqrt{X_p^2+Z_p^2}} \left(
\begin{array}{cc}
  X_p & Z_p \\
 -Z_p & X_p
\end{array}
\right) \left(
\begin{array}{c} 
x \\ 
z 
\end{array}
\right) .
\nonumber
\end{eqnarray}
Note that the integrals $W_1$ and $W_2$ are carried out over the entire
planetary disk during the complete transit phase, while the regions of
the integrals $W_3$ and $W_4$ are restricted to the overlapped region
between planetary and stellar disks during the ingress and the egress
phases.

The approximate expressions for the integrals $W_1$ and $W_2$ are
derived by expanding the integrand in powers of $\gamma$. The resultant
expressions become
\begin{eqnarray}
W_1 &\simeq &
(1-\rho^2)^{1/2}\,\,-\,\,\gamma^2\,\frac{2-\rho^2}{8(1-\rho^2)^{3/2}} +
{\cal O}(\gamma^4).
\\\
W_2 &\simeq &
(1-\rho^2)^{1/2}\,\,-\,\,\gamma^2\,\frac{4-3\rho^2}{8(1-\rho^2)^{3/2}} +
{\cal O}(\gamma^4).
\end{eqnarray}

Consider next the integrals $W_3$ and $W_4$. They correspond to the
ingress and egress phases where $\rho\simeq 1$, and the approximation
$1-\rho^2 \gg \gamma$, which is used in the perturbative expansion for
$W_1$ and $W_2$, breaks down. Thus we attempt to find better approximate
expressions as follows. Let us integrate first and rewrite $W_3$ and
$W_4$ as
\begin{eqnarray}
\label{eq:w3-new}
 W_3  &=&\frac{\pi}{6}(1-x_0)^2(2+x_0)
+\int_{x_0+\zeta-\gamma}^{x_0}d\tilde{x}~g(\tilde{x};  \eta_p, \gamma) , \\
\label{eq:w4-new}
 W_4&=&\frac{\pi}{8}(1-x_0^2)^2
+\int_{x_0+\zeta-\gamma}^{x_0}d\tilde{x}~\tilde{x}g(\tilde{x}; \eta_p,
\gamma) .
\end{eqnarray}
where the function $g$ is given by 
\begin{eqnarray}
g(\tilde{x};\,\eta_p,\,\gamma) &\equiv& (1-\tilde{x}^2)\sin^{-1}
\left\{\frac{\gamma^2-(\tilde{x}-1-\eta_p)^2}{1-\tilde{x}^2}\right\}^{1/2}
\cr
&& \quad\quad\quad 
+ \sqrt{2(1+\eta_p)(x_0-\tilde{x})\{\gamma^2-(\tilde{x}-1-\eta_p)^2\}},
\label{appen:func_g}
\end{eqnarray}
where
\begin{eqnarray}
\label{eq:x0}
 x_0 &=& 1-\frac{\gamma^2-\eta_p^2}{2(1+\eta_p)}, \\
\label{eq:zeta}
\zeta &=& \frac{2\eta_p + \gamma^2 + \eta_p^2}{2(1+\eta_p)}.
\end{eqnarray}
Equations (\ref{eq:w3-new}) and (\ref{eq:w4-new}) are still too
complicated to integrate analytically. Thus we approximate the integrand
of equation (\ref{eq:w3-new}) as
\begin{eqnarray}
g(\tilde{x};\eta_p,\gamma)=C_3\,\, g(x_c;\eta_p,\gamma)
\sqrt{\frac{(x_0-\tilde{x})(\tilde{x}-x_0-\zeta+\gamma)}
{(x_0-x_c)(x_c-x_0-\zeta+\gamma)}}, 
\end{eqnarray}
where
\begin{eqnarray}
x_c &=& x_0 + s(\zeta-\gamma).
\end{eqnarray}
In the above equations, we determine the two constants $s$ and $C_3$ as
follows. From the continuity of the first and second expressions of
equation (\ref{eq:func-A}), $C_3$ is written in terms of $s$:
\begin{eqnarray}
C_3 = \frac{4\sqrt{s(1-s)} \gamma}{g(1-2s\gamma; -\gamma,\gamma)} 
W_1(1-\gamma).
\end{eqnarray}
The value of $s$ is obtained empirically obtained to yield the best-fit
to the numerical results. Here we adopt $s=0.4$ which gives a slightly
better fit than $s=0.5$ that was used in Paper I. Finally we obtain
\begin{equation}
 W_3 \simeq \frac{\pi}{6}(1-x_0)^2(2+x_0)+\frac{\pi}{2}\gamma(\gamma-\zeta)
  \frac{g(x_c;\eta_p,\gamma)}{g(1-2s\gamma; -\gamma, \gamma)}W_1(1-\gamma).
\end{equation}

We repeat a similar procedure for equation (\ref{eq:w4-new}). In this case,
the constant $C_3$ is replaced by
\begin{eqnarray}
C_4 = \frac{4\sqrt{s(1-s)} \gamma}{g(1-2s\gamma; -\gamma,\gamma)} 
W_2(1-\gamma), 
\end{eqnarray}
and we adopt $s=0.4$ as above. Then  an approximate
analytic expression for $W_4$ is given as
\begin{equation}
 W_4 \simeq \frac{\pi}{8}(1-x_0^2)^2+\frac{\pi}{4}\gamma(\gamma-\zeta)
  (2x_0+\zeta-\gamma)
  \frac{g(x_c;\eta_p,\gamma)}{g(1-2s\gamma; -\gamma, \gamma)} W_2(1-\gamma).
\end{equation}


\begin{deluxetable}{ccl}
 \tablecaption{List of parameters
 \label{tbl:parameters}}
 \tablehead{\colhead{Variables} & \colhead{Fiducial value} & 
 \colhead{Meaning} }
 \startdata
\cutinhead{Internal parameters of star}
 $R_s$ & 1.146$R_\odot$& Stellar radius\\
 $u_1$ & 0.2925 & Limb darkening parameter\\
 $u_2$ & 0.3475 & Limb darkening parameter\\
 $\Omega_s$ & $R_s\Omega_s\sin I_s=4700\,{\rm m\,s^{-1}}$ 
& Spin rotation velocity of star\\
 $I_s$ & - & Inclination between stellar spin axis and $y$-axis\\
\cutinhead{Internal parameters of planet and ring}
 $R_p$ & 1.347$R_J$ & Planetary radius\\
 $\lambda$ & $0^\circ$ & Angle between $z$-axis \\
          & & and normal vector $\hat{\bm{n}}_p$ on $(x,z)$-plane\\
 $R_{\rm in}$ & 1.5$R_p$ & Radius of inner edge of ring\\
 $R_{\rm out}$ & 2.0$R_p$ & Radius of outer edge of ring\\
 $\tau$ & 1.0 & Optical depth normal to ring surface\\
 $\theta$ & $45^\circ$ & Obliquity angle of ring axis to $z$ axis\\
 $\phi$ & $0^\circ$ & Azimuthal angle of ring axis around $z$ axis\\
\cutinhead{Orbital parameters}
 $a$ & $3$AU & Semi-major axis\\
 $P_{\rm orbit}$ & $1809.01$days & Orbital period \\
 $i_{\rm orbit}$ & $90^{\circ}$ & Inclination angle between \\
                 &                 & normal direction
 of orbital plane and $y$-axis\\ 
$b$ & - & Impact parameter, $b=(a/R_s) \cos i_{\rm orbit}$
 \enddata
\end{deluxetable}


\end{document}